\documentclass[lettersize,journal]{IEEEtran}
\usepackage{amsmath,amsfonts}
\usepackage{array}
\usepackage{textcomp}
\usepackage{stfloats}
\usepackage{url}
\usepackage{verbatim}
\usepackage{hyperref}
\usepackage{multirow}
\usepackage{pgf,tikz}
\usepackage{enumitem}
\usepackage{float}
\usepackage{comment}
\usepackage{amsfonts}
\usepackage[ruled,vlined]{algorithm2e}
\usepackage[font=normalsize]{caption}
\usepackage{graphicx}
\usepackage{amsmath,amsthm,amssymb,mathtools,graphicx,cite,calc,color,dsfont,psfrag} 
\usepackage{array}
\usepackage{bm}
\usepackage{enumitem}
\usepackage{subfigure}
\usepackage{hyperref}
\usepackage{xcolor}
\usepackage{makecell}
\usepackage{cite}
\usepackage{array}
\usepackage{makecell}     
\usepackage{tabularx}     
\usepackage{array}        
\usepackage{booktabs}     
\newcolumntype{L}{>{\raggedright\arraybackslash}p{2.5cm}}  
\newcolumntype{C}{>{\centering\arraybackslash}p{2.5cm}}    
\newcolumntype{R}{>{\raggedleft\arraybackslash}p{2.5cm}}   

\newcolumntype{L}{>{\raggedright\arraybackslash}p{1.3cm}}  
\newcolumntype{C}{>{\centering\arraybackslash}p{1.3cm}}    
\newcolumntype{R}{>{\raggedleft\arraybackslash}p{1.3cm}}   

\usepackage[mathscr]{euscript}
\makeatletter
\newcommand{\thickhline}{%
    \noalign {\ifnum 0=`}\fi \hrule height 1pt
    \futurelet \reserved@a \@xhline
}
\newcolumntype{"}{@{\hskip\tabcolsep\vrule width 1pt\hskip\tabcolsep}}
\makeatother
\usepackage{scalerel}
\usepackage{tikz}
\usetikzlibrary{svg.path}

\definecolor{orcidlogocol}{HTML}{A6CE39}
\tikzset{
  orcidlogo/.pic={
    \fill[orcidlogocol] svg{M256,128c0,70.7-57.3,128-128,128C57.3,256,0,198.7,0,128C0,57.3,57.3,0,128,0C198.7,0,256,57.3,256,128z};
    \fill[white] svg{M86.3,186.2H70.9V79.1h15.4v48.4V186.2z}
                 svg{M108.9,79.1h41.6c39.6,0,57,28.3,57,53.6c0,27.5-21.5,53.6-56.8,53.6h-41.8V79.1z M124.3,172.4h24.5c34.9,0,42.9-26.5,42.9-39.7c0-21.5-13.7-39.7-43.7-39.7h-23.7V172.4z}
                 svg{M88.7,56.8c0,5.5-4.5,10.1-10.1,10.1c-5.6,0-10.1-4.6-10.1-10.1c0-5.6,4.5-10.1,10.1-10.1C84.2,46.7,88.7,51.3,88.7,56.8z};
  }
}

\newcommand\orcidicon[1]{\href{https://orcid.org/#1}{\mbox{\scalerel*{
\begin{tikzpicture}[yscale=-1,transform shape]
\pic{orcidlogo};
\end{tikzpicture}
}{|}}}}

\usepackage{hyperref}

\begin{document}

\title{A High-throughput and Secure Coded Blockchain for IoT}

\author{Amirhossein Taherpour \orcidicon{0000-0003-4647-102X}\,, and Xiaodong Wang \orcidicon{0000-0002-2945-9240}\,,~\IEEEmembership{Fellow,~IEEE}
\thanks{Electrical Engineering Department, Columbia University\\E-mails: at3532@columbia.edu, xw2008@columbia.edu}
}
\markboth{}%
{}

\IEEEpubid{}

\maketitle

\begin{abstract}
We propose a new coded blockchain scheme suitable for the Internet-of-Things (IoT) network. In contrast to existing works for coded blockchains, especially blockchain-of-things, the proposed scheme is more realistic, practical, and secure while achieving high throughput. This is accomplished by: 1) modeling the variety of transactions using a reward model, based on which an optimization problem is solved to select transactions that are more accessible and cheaper computational-wise to be processed together; 2) a transaction-based and lightweight consensus algorithm that emphasizes on using the minimum possible number of miners for processing the transactions; and 3) employing the raptor codes with linear-time encoding and decoding which results in requiring lower storage to maintain the blockchain and having a higher throughput. We provide detailed analysis and simulation results on the proposed scheme and compare it with the state-of-the-art coded IoT blockchain schemes including Polyshard and LCB, to show the advantages of our proposed scheme in terms of security, storage, decentralization, and throughput.

\end{abstract}
\section{Introduction}
\IEEEPARstart{T}{he} Internet of Things (IoT) has become one of the most important technologies in the 21st century, which conceptualizes a hyperconnected world where physical things (ranging from small embedded sensors and wearable gadgets to connected vehicles and automated systems) can share and collect data with minimal human intervention. IoT facilitates accomplishing various tasks in people's daily lives and enhances their interactions with their surroundings. However, the increases in size, complexity, and the number of connected devices in IoT networks pose significant challenges in a number of aspects including centralization, authenticity, reliability, anonymity, and security~\cite{R1,R2,R3}.\par
Blockchain technology which is a distributed immutable ledger with high transparency and reliability has received considerable attention as one of the promising solutions to some of the above challenges with IoT networks~\cite{R4,R5,R6}. Although the blockchain has a number of advantages, the main issue is that finding a proper consensus algorithm for IoT blockchains is challenging since existing consensus algorithms for conventional blockchains, e.g., Proof of Work (PoW), Proof of Stake (PoS), and some others that were proposed later, e.g., practical Byzantine fault tolerance (pBFT), Proof of Authority (PoA) do not meet the stringent requirements of IoT blockchains, i.e., real-time transaction settlement and high security ~\cite{R7,R8,R9,R10,Rk1,Rk2}. Since an IoT blockchain consists of many heterogeneous devices with diverse functionalities, the network could produce millions of transactions in a short period of time. At the same time, some of them have to be processed quickly with the highest accuracy possible. The above-mentioned existing consensus algorithms either do not have the desired high throughput or can achieve the throughput only by sacrificing the network security. Another problem with the existing blockchains is that all data are replicated in each node and therefore storage becomes a major problem after some time. This issue is more severe with IoT blockchains with millions of transactions and many resource-restricted devices and some of which might be miners as well~\cite{R11}.\par
Recently coded blockchain techniques have been proposed ~\cite{R4,R11,R12,R13,R14,R15} to achieve better throughput and security along with reduced size of the blockchain. Some of these works were inspired by coded distributed computing (CDC) ~\cite{R100,R101,R102,R16,R17,R18}. The basic idea in CDC is that when a computation task needs to be done, working on encoded data provides the opportunity to divide the whole computation into sub-tasks, each of which could be done independently and a subset of the computation results of those sub-tasks suffice to obtain the final result of the original computational task.\par
In particular, Polyshard~\cite{R14} makes use of the notion of sharding in blockchain~\cite{R22,R23,R24,R25,R26} and employs the Lagrange Coded Computing scheme (LCC)~\cite{R27}. A sharded blockchain splits a blockchain network into smaller partitions, known as ``shards." Each keeps its own data, which is distinctive and independent from data in other shards; and the nodes in each shard are only responsible for managing transactions in their shard. In Polyshard, to verify blocks generated within a shard, nodes individually compute a polynomial verification function over a coded chain to find an intermediate result, and having intermediate results from all other miners in the network, nodes can reach a consensus about the validity of each block. However, Polyshard suffers from many problems such as, delay, security, and oversimplification of the model ~\cite{R4,R28,R29,R30,R31}. ~\cite{R29} tries to make the Polyshard more practical by considering a cross-sharded blockchain where nodes that are not in the same shard still can place a transaction in the network, a crucial feature that Polyshard lacks. The Lagrange Coded Blockchain (LCB)~\cite{R4} is a non-sharded approach that tries to improve Polyshard by using the LCC in both block generation and verification stages. Also, to solve the security issue of Polyshard that arises when a group of miners, called stragglers, are not capable of sending the computation results in time, LCB assigns more than one computation task to miners that are faster and capable of doing more computations. However, the LCB scheme is not storage efficient, which is a significant drawback for IoT networks. Furthermore, it gives rise to some other security issues and leads to a centralized network which will be discussed in this paper. While~\cite{R12} provides a secure naming and storage system using blockchain technology, it is primarily focused on static data storage and naming services, making it unsuitable for high-frequency, real-time transaction processing typical in IoT networks. Moreover, Blockstack's design in  ~\cite{R12} does not inherently address the scalability and storage efficiency challenges posed by IoT devices with limited resources, nor does it incorporate advanced coding or sharding techniques to optimize throughput and minimize latency in dynamic IoT environments. In~\cite{R11,R13}, besides proposing a storage efficient blockchain, authors aim to minimize the bootstrapping cost using some coding techniques, while throughput-wise, these schemes do not scale well since these works employ the conventional consensus algorithms like PoW. \par
From the above discussions, each of the existing works in coded blockchain has some drawbacks. Particularly, when contemplating the formulation of an IoT blockchain framework, certain crucial aspects necessitate meticulous deliberation, thereby unveiling the conspicuous inadequacies prevalent in the aforementioned studies:  
\begin{enumerate}
    \item Enhanced Efficiency: The existing blockchains are inefficient in handling the high volume of transactions generated in IoT networks. By designing a new IoT blockchain, we can significantly increase throughput and overcome the limitations of existing systems, ensuring seamless operation and efficient processing of IoT transactions.
    \item Priority-based Transaction Handling: IoT applications require different levels of transaction priority based on their versatility and importance. A new IoT blockchain can incorporate sophisticated transaction selection and consensus procedures to ensure that critical transactions receive the appropriate priority, enhancing the overall functionality and effectiveness of IoT applications.
    \item Overcoming Scalability, Resource Constraints, and Latency: Scaling a blockchain to handle the high transaction throughput and storage requirements of IoT data, accommodating resource-constrained devices, and providing low-latency communication are significant challenges. A new IoT blockchain can address these challenges through innovative solutions, enabling robust scalability, efficient operation on resource-constrained devices, and quick transaction confirmation for real-time IoT applications.
\end{enumerate}
In this paper, we propose a new coded blockchain for IoT networks that does not suffer from any of the drawbacks mentioned in existing works and considers the above aspects. Here we list the main contributions of this article:
\begin{enumerate}
\item We introduce the concept of transaction selection in the context of an IoT network to prioritize different types of transactions. By formulating the priority of transactions and addressing it as a stochastic or deterministic optimization problem, we can identify and select transactions of high priority for processing achieving high throughput in the system.
\item Considering a dynamic IoT network in which nodes may join and leave the network, suitable for IoT scenario, we propose a lightweight consensus algorithm that along with the transaction prioritization and the low-complexity decoding procedure for fetching the required data for verification, help to reduce the futile work and result in a high throughput.
\item We discuss in-depth the notions of security, storage reduction, decentralization, and throughput in coded blockchains and compare our proposed scheme with Polyshard and LCB in terms of these metrics.
\end{enumerate}
The remainder of the paper is organized as follows. Section II introduces the blockchain of things system model and the existing coded blockchains. Section III is an overview of our proposed scheme. In Section IV, we provide details on the transaction selection and assignment stage of our proposed scheme. In Section V, through simulations we compare our scheme with Polyshard and LCB schemes in terms of storage reduction, decentralization, throughput, and discuss the security issues. Finally, Section VI concludes the paper.

\section{System Descriptions and Background} A typical IoT system consists of three elements: 1) IoT devices that produce data. These devices can be in a wide variety ~\cite{R50}, including industrial sensors working with Low Power Wide Area Networks (LPWANs), smart retail that uses Bluetooth Low-Energy (BLE), logistics and asset tracking operating with Radio Frequency Identification (RFID), wearables, and smart homes deploying cellular/Wi-Fi connectivity. 2) Base station (BS), which receives the data from IoT devices and aggregates them to form batches that are sent to miners. 3) Miners that verify the validity of given data from the BS, and after reaching a consensus, the validated data are appended to the blockchain in the form of blocks. Next, we explain the mining process in an IoT blockchain in detail.\par
\subsection{Overview of Coded vs. Uncoded Blockchain Architectures}
Blockchains can be fundamentally classified into \textit{uncoded} and \textit{coded} blockchains~\cite{Revised1}. This classification is pivotal for understanding the subsequent sections introduced in this paper.\par
\subsubsection{Uncoded Blockchain}
In an \emph{uncoded} blockchain architecture, each data block is fully replicated across all network nodes, impacting both storage and retrieval processes.\par

\textbf{Storage.} In a traditional uncoded blockchain, each data block \( B(m) \) is fully replicated and stored by every node within the network. Specifically, for a blockchain consisting of \( t \) blocks, each node maintains the complete set \( \mathbf{B}^{t} = \{ B(1), B(2), \dots, B(t) \} \). This full replication ensures that every node holds an independent copy of all blocks on the chain. While this approach guarantees data availability, it incurs substantial storage overhead as the blockchain scales. Prominent blockchains, such as Bitcoin and Ethereum, exemplify this limitation, requiring nodes to store hundreds of gigabytes of data, a demand that escalates as new blocks are added, posing significant scalability challenges.\par

\textbf{Recovery and Verification.} Recovery and verification are straightforward in an uncoded blockchain. When a specific block \( B(i) \) is required, each node accesses it directly from its local storage. This process allows for efficient retrieval speed and enables verification without relying on other nodes. For example, to validate a transaction within block \( B(i) \), a node retrieves \( B(i) \) directly from its local copy \( \mathbf{B}^{t} \), eliminating the need for external assistance. However, this efficiency in retrieval comes at the cost of increased storage and communication overhead, as every node must replicate the entire blockchain, leading to significant data redundancy.\par

\subsubsection{Coded Blockchain}
In a \emph{coded} blockchain architecture, each block is divided into smaller fragments and encoded using error correction techniques, distributing these coded fragments across network nodes and providing enhanced storage efficiency and resilience.\cite{Revised2}\par

\textbf{Storage.} In a coded blockchain, each block \( B(m) \) is divided into \( k \) original fragments, denoted as \( \{ B(m)_1, B(m)_2, \ldots, B(m)_k \} \). These original fragments are then encoded to produce \( N \) coded fragments, denoted \( \tilde{B}_1(m), \tilde{B}_2(m), \dots, \tilde{B}_N(m) \), which are distributed across the network nodes. Each node in the network stores only \emph{one coded fragment} of each block, so a node \( j \) would store \( \tilde{B}_j(m) \) for block \( B(m) \) in epoch \( m \).\par
These \( N \) coded fragments are constructed as \emph{linear combinations} of the \( k \) original fragments. Specifically, each coded fragment \( \tilde{B}_j(m) \) for node \( j \) is formed by linearly combining the original fragments \( \{ B(m)_1, B(m)_2, \ldots, B(m)_k \} \) with predetermined coding coefficients. Therefore, node \( j=1,\ldots,N \) stores \( \tilde{B}_j(m) \), where each \( \tilde{B}_j(m) \) is a coded fragment that can collectively reconstruct \( B(m) \) when any subset of \( k \) fragments is available.\par
This approach significantly reduces the storage burden on each node, as each node only needs to retain a fraction \( \frac{1}{N} \) of the blockchain data, improving scalability and efficiency over traditional uncoded blockchain systems.\par
\textbf{Recovery and Verification.} In a coded blockchain, recovery of a block \( B(m) \) relies on the distributed coded fragments. To reconstruct the original block for purposes such as transaction validation or consensus, any node can collect a subset of \( k \) coded fragments from the network’s \( N \) nodes. Using a decoding algorithm, such as Reed-Solomon decoding, the node combines these \( k \) coded fragments to reconstruct \( B(m) \) fully. The linear combination encoding provides robustness, as any \( k \) out of \( N \) coded fragments are sufficient for complete recovery, ensuring data availability even if some nodes fail or their data becomes corrupted.\par
The security parameters of the consensus protocol and the encoding scheme parameters (\(k, N\)) are conceptually distinct yet indirectly interconnected through system redundancy and data recovery capabilities. The consensus mechanism ensures correct decision-making even in the presence of faulty or adversarial nodes, while the encoding scheme optimizes storage efficiency and data availability via its chosen \(k\) and \(N\) values. Although the consensus fault tolerance does not directly dictate encoding choices, selecting a higher \(N/k\) ratio enhances data redundancy, which in turn bolsters fault tolerance by safeguarding against data loss or corruption amid node failures or malicious activity. This enhanced redundancy ensures reliable reconstruction of original data, thus supporting the accuracy and integrity of consensus outcomes. However, these benefits come with increased storage, communication, and computational overhead, necessitating a balanced, coordinated tuning of \(k\) and \(N\). Such tuning aligns encoding parameters with security requirements, achieving the desired level of resilience without compromising system efficiency.\par
Table~\ref{tab:comparison} provides a comprehensive comparison between coded and uncoded blockchains, highlighting their distinct characteristics in terms of storage mechanisms, scalability, fault tolerance, data retrieval, redundancy, communication overhead, and implementation complexity\cite{Revised1}. This comparison underscores the coded blockchain's suitability for environments with constrained storage and bandwidth, such as IoT networks, by addressing the storage and communication inefficiencies that limit the scalability of conventional blockchain systems.

\begin{table*}[!t] 
    \centering
    \caption{Comparison Between Coded and Uncoded Blockchains}
    \label{tab:comparison}
    \small
    \renewcommand{\arraystretch}{1.2} 
    \resizebox{\textwidth}{!}{ 
        \begin{tabular}{|>{\centering\arraybackslash}p{3.2cm}|>{\centering\arraybackslash}p{6cm}|>{\centering\arraybackslash}p{6cm}|}
            \hline
            \textbf{Aspect} & \textbf{Uncoded Blockchain} & \textbf{Coded Blockchain} \\
            \hline
            Storage Mechanism & 
            Full replication of all blocks by each node. & 
            Storage of encoded fragments; only a subset is required for recovery. \\
            \hline
            Scalability & 
            Limited by linear growth in storage and computational needs. & 
            Enhanced scalability through distributed storage and processing. \\
            \hline
            Fault Tolerance & 
            Vulnerable to node failures that affect data availability. & 
            Resilient to node failures, tolerating up to \( n - k \) faults. \\
            \hline
            Data Retrieval & 
            Direct access to complete blocks by each node. & 
            Requires decoding from any \( k \) fragments to reconstruct a block. \\
            \hline
            Redundancy & 
            High redundancy; each node stores the entire blockchain. & 
            Lower redundancy; each node holds only a fraction of the blockchain. \\
            \hline
            Communication Overhead & 
            Higher bandwidth requirements due to full block transmission. & 
            Reduced bandwidth usage with coded fragment transmission. \\
            \hline
            Implementation Complexity & 
            Lower complexity due to straightforward data storage. & 
            Higher complexity due to encoding and decoding mechanisms. \\
            \hline
        \end{tabular}
    }
\end{table*}

\subsection{Mining Process in IoT Uncoded Blockchains} We divide the blockchain time-wise into epochs and assume that one block is appended to the blockchain after each epoch. Suppose, at the beginning of epoch $t$, the system consists of $N(t)$ miners. One epoch of the mining process starts with collecting the data from all devices connected to the BS. Denote the matrix of data collected by the BS by $T(t)=[L(t) \quad S(t) \quad R(t)]$ that consists of $n$ transactions, each corresponding to one row. Here, \( L(t) \) encompasses metadata such as transaction ID, sender, recipient, amount, and contract type; \( S(t) \) represents state information detailing contract conditions and post-execution updates; and \( R(t) \) specifies the functions invoked for contract execution.\par 
Matrix-based representations of transactions and blocks are grounded in extensive research within the coded blockchain domain, demonstrating significant improvements in scalability, storage efficiency, and computational performance~\cite{R11,R5000, Revised3, Revised4}. Building upon these established methodologies, our approach tailors matrix structures specifically to address the constraints and requirements of IoT environments.\par

In our framework, transactions are organized as rows within a transaction matrix \( T(t) = [L(t) \quad S(t) \quad R(t)] \). Similarly, blocks are structured as columns within a block matrix \( B(m) = [L^*(m) \quad S^*(m) \quad R^*(m)] \), aggregating validated transactions and states for each epoch \( m \). This organization facilitates efficient encoding and storage across network nodes.\par
To illustrate, consider Ethereum blockchain where the metadata \( L(t) \) includes transaction hash, sender and recipient addresses, value transferred, transaction fees, and contract type. The state information \( S(t) \) captures the contract's current state, such as balances, contract-specific variables, and execution conditions, while \( R(t) \) details smart contract functions invoked, such as \texttt{transfer}, \texttt{approve}, or \texttt{notify}. This alignment ensures compatibility with existing blockchain frameworks, leveraging well-understood structures to enhance our coded blockchain's functionality.\par

\begin{table*}[ht]
\centering
\renewcommand{\arraystretch}{1.1}
\caption{Matrix Representation of a Block with Grouped Columns for \( L \) (Metadata), \( S \) (State Information), and \( R \) (Functions)}
\label{tab:block_matrix}
\resizebox{\textwidth}{!}{%
\begin{tabular}{|l|l l l l|l l l|l l l l|}
\hline
\textbf{Transaction} & \multicolumn{4}{c|}{\textbf{\( L \): Metadata}} & \multicolumn{3}{c|}{\textbf{\( S \): State Information}} & \multicolumn{4}{c|}{\textbf{\( R \): Functions}} \\ \hline
 & \textbf{Transaction ID} & \textbf{Sender} & \textbf{Recipient} & \textbf{Amount} & \textbf{Contract} & \textbf{Status} & \textbf{Block Number} & \textbf{Function} & \textbf{Gas} & \textbf{Event} & \textbf{Callback} \\ \hline
\textbf{Transaction 1} & 0xabc & 0xS1 & 0xR1 & 2 & 0xC1 & Confirmed & 1250 & transfer & 21000 & Transfer & UpdateBalance \\ \hline
\textbf{Transaction 2} & 0xdef & 0xS2 & 0xR2 & 3 & 0xC2 & Pending & 1251 & approve & 30000 & Approval & InitiatePayment \\ \hline
\textbf{Transaction 3} & 0xghi & 0xS3 & 0xR3 & 1 & 0xC3 & Executed & 1252 & notify & 25000 & Notify & AlertDoctor \\ \hline
\end{tabular}%
}
\end{table*}

Table~\ref{tab:block_matrix} presents a concrete matrix representation of transactions, where each row corresponds to a transaction segmented into Metadata (\( L \)), State Information (\( S \)), and Functions (\( R \)). Specifically, the metadata includes Transaction ID, Sender, Recipient, Amount, and Contract type; the state information captures Contract address, Status (e.g., Pending, Confirmed, Executed), and Block Number; and the functions specify the invoked function (e.g., \texttt{transfer}, \texttt{approve}, \texttt{notify}), Gas allocated, Event triggered, and Callback functions. This structured representation aligns with Ethereum's blockchain, facilitating efficient encoding, storage, and verification processes essential for our coded blockchain framework.\par

This matrix-based representation enables encoding over finite fields $\mathbb{F}_q$, allowing the application of error-correcting codes that enhance data resilience against loss and attacks, critical for decentralized coded loT networks. Matrix operations also support parallel processing of transactions, which accelerates verification and consensus through optimized linear algebra techniques, increasing system throughput and benefiting loT networks with limited computational resources. Encoding blocks as matrices enhances resilience against tampering, ensuring data integrity and supporting robust security protocols. This systematic coding approach further obscures transaction details, adding an additional layer of security\par

We next illustrate these three components of the transaction data using a real-life application.\par
Consider remote patient monitoring by which, with the help of wearable medical devices, healthcare workers can monitor and capture medical and other health data from patients for assessment and, when necessary, treatment. A physician deploys a contract that includes a function ``notify." Let the time of deployment be $t^{\prime}<t$, so that this contract appears as a row in the matrix $T(t^{\prime})=[L(t^{\prime}) \quad S(t^{\prime}) \quad R(t^{\prime})]$, where $L(t^{\prime})$ includes the patient's information consisting of the patient history, information that results in the uniqueness of the device that is being used for monitoring the patient status, public keys that restrict the access to the contract to a particular group of people like trusted doctors and patient's family, etc. $S(t^{\prime})$ contains pre-defined conditions that when met, the functions in $R(t^{\prime})$ including the function ``notify," must be called. While the wearable devices keep monitoring the patient's health status, when a device detects an abnormality, it submits a transaction to the network, which corresponds to one row in matrix $T(t)=[L(t) \quad S(t) \quad R(t)]$. Here $L(t)$ calls the deployed contract at epoch $t^{\prime}$ that corresponds to this patient. $S(t)$ states the patient's present status and that she needs to be hospitalized or be notified. $R(t)$ contains the functions that need to be called for this purpose. It is the responsibility of the miners in the network to check the validity of the information in $L(t)$ and $S(t)$ with the corresponding existing information in the blockchain, which in our example is the related rows in $L(t^{\prime})$ and $S(t^{\prime})$. In case of validity, the ``notify" function is executed, letting the doctors know about the mentioned patient. The state $S(t)$ will be updated and recorded in the blockchain.\par
After forming matrix $T(t)$, the BS sends this matrix to all miners in the network. Each miner checks the validity of the $n$ contracts in $T(t)$ based on the existing stored ledger in the previous blocks, i.e., $\mathbf{B}^{t-1}=\{B(1), \ldots, B(t-1)\}$ where $B(m)=[L^{*}(m) \quad S^{*}(m) \quad R^{*}(m)]$ consists of validated rows of $L(m)$, $S(m)$, and $R(m)$, respectively. To check the validity of the contracts in $T(t)$, miner $j$ computes $v_{j}(t)=V_{j}^{t}(L(t), S(t),\mathbf{B}^{t-1})\in \{0,1\}^{n}$, where $V_{j}^{t}$ is the verification operator for miner $j$ at epoch $t$, and 1/0 indicate whether or not a contract is valid. Next, each miner sends its verification result vector to other miners. All miners in the network apply a consensus algorithm to determine the global validity of the contracts by computing $C(\left\{v_{j}(t)\right\}_{i=1}^{N})=v^{*}(t)\in \{0,1\}^{n}$. For instance, under the majority rule, $C$ selects the set of transactions that are declared valid by the majority of miners. Then, using $v^{*}(t)$, $L^{*}(t)$ and $R^{*}(t)$ can be determined.\par
Each miner $j$ executes the valid contracts by calling the functions in $R^{*}(t)$, i.e., it computes $F_{j}^{t}(R^{*}(t),S(t))=S_{j}(t)$, where $F_{j}^{t}$ is the function calling operator for miner $j$, that updates the state of the valid contracts to $S_{j}(t)$, which is then broadcasted to the network. Given the updated states from all miners, the confirmed updated state $S^{*}(t)$ is obtained by computing $U(\left\{S_{j}^{t}\right\}_{j=1}^{N(t)})=S^{*}(t)$, where similar to $C$, $U$ is an operator that is applied to the received updated states from all miners, and determines the updated states of transactions that are in consensus. The updated state $S^{*}(t)$ along with $L^{*}(t)$ and $R^{*}(t)$ forms the block $B(t)=[L^{*}(t) \quad S^{*}(t) \quad R^{*}(t)]$ for epoch $t$. Finally, each miner appends $B(t)$ as the new block to its local copy of the blockchain $\mathbf{B}^{t}={\bf B}^{t-1} \cup B(t)$.\par
The above conventional blockchain suffers from the so-called trilemma problem ~\cite{R19,R20,R21}, i.e., the problem of being unable to balance simultaneously security, decentralization, and scalability, which means when one or two of these three metrics improve, the rest will be compromised. Coding techniques are one of the promising approaches to addressing the trilemma problem in the blockchain. Next, we will explain the mining process in the existing coded IoT blockchain.
\subsection{Features of LCC-based Coded Blockchains} In addition to having low throughput, another drawback of the conventional consensus algorithms is that a small fraction of miners make the majority of network decisions~\cite{R43,R51,R52}. In a coded blockchain, the goal is to have a decentralized high throughput scheme where all miners work together on computational tasks. We focus on the coded blockchain schemes LCB ~\cite{R4} and Polyshard~\cite{R14} that employ the Lagrange Coded Computing (LCC). Although the two schemes apply the LCC in different ways, they share common features that are described below.\par
Suppose that the network wants to reach consensus about the validity of the data block $X(t)$. Each miner $j$ first obtains the encoded block $\tilde{X}_{j}(t)=E_{j}^{t}(X(t))$, where $E_{j}^{t}$ is the encoding operator for miner $j$ which is a Lagrange interpolator at a specific point. Meanwhile, in a coded blockchain, instead of the raw data, miners store the encoded blocks of the previous epochs. Therefore, miner $j$ has access to $\tilde{\mathbf{B}}_{j}^{t-1}=\{\tilde{B}_{j}(1), \ldots, \tilde{B}_{j}(t-1)\}$ which is its share of the coded blocks. Then miner $j$ applies the operator $H_{j}^{t}$ to the received data and computes $H_{j}^{t}(\tilde{X}_{j}(t),\tilde{\mathbf{B}}_{j}^{t-1})=\tilde{h}_{j}^{t}$ and broadcasts the results to all other miners, where $H_{j}^{t}$ is a verification function on coded data. Note that unlike the uncoded blockchain, $\tilde{h}_{j}^{t}$ is not the verification results of miner $j$, but an intermediate coded result. Having the intermediate results from all miners, each miner $j$ computes $D^{t}(\{\tilde{h}_{i}^{t}\}_{i=1}^{N(t)})$, where $D^{t}$ is the decoding operator, to find the consensus decision on the validity of $X(t)$. An important feature of the LCC scheme is that miners can know the consensus decision even if some of the $\tilde{h}_{i}^{t}$'s are erroneous, i.e., the $i$-th miner is malicious, or are not received, i.e., the $i$-th miner is a straggler. Therefore, miners just need to wait for the intermediate results from a subset of the fastest honest miners to reach the consensus decision. Another feature is that miners store the ledger in coded form, and that the computation is performed on coded data. It should be noted that in Polyshard~\cite{R14} LCC is used for the block verification stage only, while in LCB ~\cite{R4} LCC is used for both block generation and verification stages.\par

\subsection{Challenges in Uncoded and Coded IoT Blockchains} First, the number of transactions generated in IoT networks is typically significantly higher than that in conventional blockchains. The low throughput of conventional blockchains makes them highly inefficient. On the other hand, in the existing coded schemes, the Collateral Invalidation (CI) rate~\cite{R29}, which is the number of abandoned transactions due to one invalid transaction, is very high, making them inefficient for IoT blockchains.\par
Second, in the existing blockchains, the mechanism for faster processing is the higher transaction fees. However, in IoT applications, all transactions do not have the same priority due to their versatility. For example, transactions that control a health center or security of a sensitive place have higher priority than those in retail environments for automatic checkout. Therefore, the transaction selection and consensus procedure must be designed more carefully.\par

\section{Proposed Coded IoT Blockchain} To address the above-mentioned shortcomings of the existing coded IoT blockchains, and to simultaneously achieve a high level of security, decentralization, and scalability, we propose a new coded IoT blockchain that is based on our recent work of rateless coded blockchain~\cite{R5000}. The main new component of the proposed scheme is a transaction assignment stage designed to increase the throughput. In this section, we describe the rateless coded blockchain structure and the main steps of the mining procedure of the proposed coded IoT blockchain. Sections V will compare our proposed system with the existing coded IoT blockchains under different performance metrics.\par
\subsection{Problem Statement}
Throughput (\(\Theta(t)\)) in blockchain quantifies the number of transactions processed and appended to the blockchain per epoch—a fixed interval during which transactions are collected, validated, and stored. Maximizing throughput involves selecting an optimal subset of transactions \( \mathcal{T}_s(t) \subseteq \mathcal{T}(t) \) from the incoming pool \( \mathcal{T}(t) \), ensuring that resource constraints are met while minimizing computational waste. This optimization problem is expressed as:

\begin{align}\label{first_throughput}
\text{Maximize} \quad & \Theta(t) = \sum_{j \in \mathcal{T}_s(t)} \mathbf{1}_{\operatorname{valid}(j)}, \\
\text{s.t.} \quad & \sum_{j \in \mathcal{T}_s(t)} \mathcal{P}_j \leq \mathcal{R}(t),
\end{align}

where \(\mathbf{1}_{\text{valid}(j)} = 1\) if transaction \(j\) is validated and successfully appended to the blockchain as part of a block, and \(\mathbf{1}_{\text{valid}(j)} = 0\) otherwise. Here, \( \mathcal{P}_j \) represents transaction \( j \)’s resource consumption, and \( \mathcal{R}(t) \) is the resource capacity available during the epoch.

To enhance throughput optimization, three key system constraints are defined in this paper. The block size constraint ensures the total size of selected transactions does not exceed the encoded block’s capacity (\(\mathcal{R}^{\text{size}}\)), defined as \( \sum_{j \in \mathcal{T}_s(t)} \mathcal{P}_j^{\text{size}} \leq \mathcal{R}^{\text{size}} \). This balances block utilization with storage efficiency. The computational capacity constraint governs the network’s ability to encode, decode, and validate transactions, ensuring that \( \sum_{j \in \mathcal{T}_s(t)} \mathcal{P}_j^{\text{comp}} \leq \mathcal{R}^{\text{comp}} \). This prevents processing delays and overloads, maintaining operational throughput despite the added complexity of coded data.\par
Additionally, validation depth constraints address the overhead introduced by accessing historical encoded data for transaction validation. Transactions with greater validation depth (\(\mathcal{P}_j^{\text{depth}}\)) require more intensive decoding operations to retrieve relevant blocks that are buried deeper in blockchain. The system enforces a maximum validation depth (\(\mathcal{R}^{\text{depth}}\)) to prioritize transactions with lower validation requirements, reducing computational burden and preserving system efficiency as setting $\max_{j \in \mathcal{T}_s(t)} \mathcal{P}_j^{\text{depth}} \leq \mathcal{R}^{\text{depth}}$ ensuring encoded blockchain systems maintain high throughput despite the complexity of data retrieval.\par
Moreover, to transcend the goal of merely maximizing the quantity of processed transactions, the proposed framework incorporates a \textbf{transaction prioritization} mechanism. This mechanism shifts the objective from solely maximizing the number of transactions to maximizing the \textbf{cumulative reward}, denoted by \( \Theta_{\text{eff}}(t) \), which accounts for the intrinsic value of each transaction based on factors such as urgency, fees, and age. Formally, the enhanced optimization problem considering the three key system constraints is defined as:
\begin{align}\label{second_throughput}
\text{Maximize} \quad & \Theta_{\text{eff}}(t) = \sum_{j \in \mathcal{T}_s(t)} r_j\mathbf{1}_{\operatorname{valid}(j)}, \\
\text{s.t.} \quad & \sum_{j \in \mathcal{T}_s(t)} \mathcal{P}_j^{\text{size}} \leq \mathcal{R}^{\text{size}} \label{constraint:size}, \\
& \sum_{j \in \mathcal{T}_s(t)} \mathcal{P}_j^{\text{comp}} \leq \mathcal{R}^{\text{comp}} \label{constraint:comp}, \\
& \max_{j \in \mathcal{T}_s(t)} \mathcal{P}_j^{\text{depth}} \leq \mathcal{R}^{\text{depth}} \label{constraint:depth}.
\end{align}
where in this refined formulation, \(r_j\) denotes the reward associated with transaction \( j\), which reflects its utility based on vitality, transaction fees, and age.\par
Furthermore, efficient miner allocation enhances throughput by distributing computational tasks effectively, reducing validation time and delays, and optimizing resource use. This increases transaction validation capacity and enables the network to handle high volumes without compromising performance.\par
In the following, in Section IV-A, we discuss how urgency, fees, and age of transactions are incorporated into \( \Theta_{\text{eff}}(t) \) as a reward maximization problem, followed by modeling problem (\ref{second_throughput})-(\ref{constraint:depth}) as a combinatorial optimization problem. Then, in Section IV-B, we address determining the optimal number of miners for processing the transactions.\par

\subsection{Background on Rateless Coded Blockchain} First, we give a brief description of the encoding process employed in our blockchain. Recall that at the end of epoch $t$ the network forms block $B(t)=[L^{*}(t) \quad S^{*}(t) \quad R^{*}(t)]$ which consists of the verified rows of the matrix of received transactions $T(t)=[L(t) \quad S(t) \quad R(t)]$. Encoding is performed every $W$ epochs on $W$ consecutive blocks. That is, at epoch $t=\ell W$, each miner $j$ encodes blocks $B((\ell-1)W+1), \ldots, B(\ell W)$ into a coded block $\mathbf{c}_{j}(\ell)$ as follows. First, we represent each data block $B((\ell-1)W+w)$ with a column vector $\mathbf{b}_{\ell}(w)$ of size $s$ over the finite field $\mathbb{F}_{q}$, where $q=2^{p}$ and $p$ is the number of bits in each elements of $\mathbf{b}_{\ell}(w), w=1, \ldots, W$. Then, a systematic block code with generator $\mathbf{G}$ of size $W \times \overline{W}$ is applied to obtain $[\mathbf{b}_{\ell}(1), \ldots, \mathbf{b}_{\ell}(W)]\mathbf{G}=[\mathbf{d}_{\ell}(1), \ldots, \mathbf{d}_{\ell}(\overline{W})]$, where $\overline{W}>W$ and $\left\{\mathbf{d}_{\ell}(w),w=1, \ldots, \overline{W} \right\}$ are intermediate codewords or pre-codewords. Since $\mathbf{G}$ is systematic, we have $\mathbf{d}_{\ell}(i)=\mathbf{b}_{\ell}(i)$ for $i=1, \ldots, W$. Next each miner $j$ further encodes the $\overline{W}$ intermediate codewords into a coded block $\mathbf{c}_{\ell}(j)$ using a systematic LT code. Specifically, for $j=1, \ldots, \overline{W}$, miner $j$ simply stores $\mathbf{d}_{\ell}(j)=\mathbf{c}_{\ell}(j)$, corresponding to the information part of the code. All other miners store the parity part of the code. In particular, each miner $j$, $j=\overline{W}+1, \ldots, N(\ell W)$, first chooses a degree $L_{j}$ from a generator degree distribution $\Omega(L)$. It then randomly chooses $L_{j}$ intermediate codewords $\left\{\mathbf{d}_{\ell}(i_{1}), \ldots, \mathbf{d}_{\ell}(i_{L_{j}})\right\}$ from $\left\{\mathbf{d}_{\ell}(1), \ldots, \mathbf{d}_{\ell}(\overline{W})\right\}$ and computes $\mathbf{c}_{\ell}(j)= \mathbf{d}_{\ell}(i_{1}) \oplus \ldots \oplus \mathbf{d}_{\ell}(i_{L_{j}})$. The coded block $\mathbf{c}_{\ell}(j)$ is stored by miner $j$. Recall that $N(\ell W)$ is the number of miners in the network at the time of encoding $t=\ell W$. We choose $\overline{W}$ such that $N(\ell W)>\overline{W}>W$. The encoding scheme described above is shown in Fig. 1.\par
In our scheme miners process transactions using the raw data. Suppose miner $j$ needs access to a subset of blocks, denoted by $\mathbf{\hat{B}}_{j}(\ell)$, from the group of blocks $\left\{B((\ell-1)W+1), \ldots, B(\ell W)\right\}$, i.e., some of the systematic code words in $\left\{\mathbf{b}_{\ell}(1), \ldots, \mathbf{b}_{\ell}(W)\right\}$, $\left\{\mathbf{d}_{\ell}(1), \ldots, \mathbf{d}_{\ell}(W)\right\}$, or $\left\{\mathbf{c}_{\ell}(1), \ldots, \mathbf{c}_{\ell}(W)\right\}$. Now, to obtain the required data, there are two approaches. First, miner $j$ can repair the corresponding intermediate codewords using the RNM algorithm, which is  described in detail in Appendix, by exploiting the bitwise xor relationship between $\mathbf{c}_{\ell}(i)$'s and $\mathbf{d}_{\ell}(i)$'s. This essentially corresponds to decoding the outer LT codes. But instead of decoding all $\overline{W}$ intermediate codewords, the RNM algorithm decodes only the required intermediate blocks. Second, in case the first approach cannot obtain all required blocks, then a full decoding of both the outer LT code and inner block code needs to be carried out to obtain the group of original data blocks $\left\{\mathbf{b}_{\ell}(1), \ldots, \mathbf{b}_{\ell}(W)\right\}$. Note that for repairing, the number of involved blocks is typically proportional to the generated degree $L_{j}$, while for decoding this numb becomes proportional to the number of miners $N(\ell W)$. Since $N(\ell W)\gg L_{j}$, repairing is computationally much more efficient than decoding. Fortunately, most of the time only repairing is needed for a miner to obtain its required data and decoding is rarely needed. The details of the RNM algorithm and the decoding procedure can be found in ~\cite{R5000,R4000} and references therein.\par
Finally, we briefly discuss the choice of the size $W$ of the block group. On the one hand, since for every $W$ data block, each miner stores only one coded block, $W$ represents the compression ratio and should be large. On the other hand, if $W$ is too large and some miners may leave the network, this could result in unsuccessful decoding, and miners may not be able to recover some data blocks during the time interval between encoding instants $(\ell-1)W$ and $\ell W$. Therefore, there is a trade-off between the decoding failure probability and the reduction in storage. The method for finding the optimum $W$ is discussed in ~\cite{R5000} by using analysis tools in~\cite{R4000,R4001}.\par
\begin{figure}[!htb]
\centering
\includegraphics[width=3in]{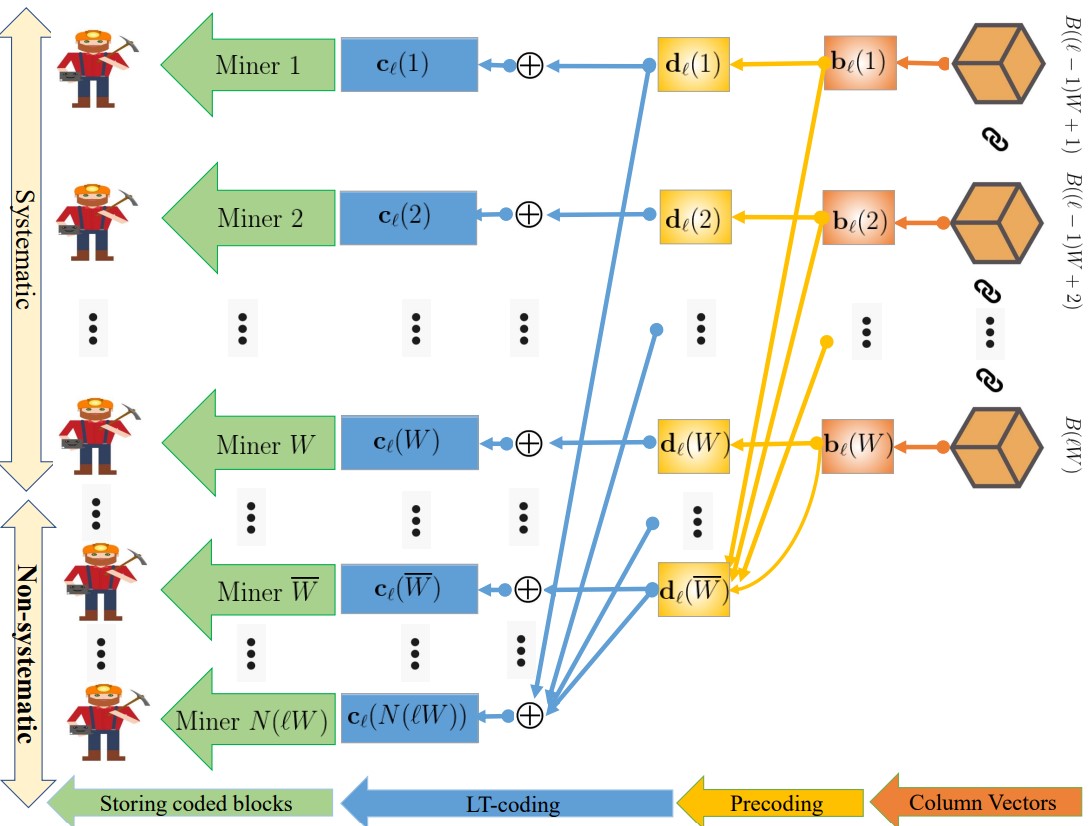}
\caption{Block coding in the proposed scheme using a systematic raptor code.}
\label{thglrd31}
\end{figure}

\subsection{Overview of the Mining Process}
The mining process in our approach goes beyond the conventional task of merely generating blocks; it’s a carefully structured, multi-stage mechanism tailored to the unique demands of IoT blockchains. Traditional mining typically involves validating transactions or solving cryptographic puzzles to create blocks. However, IoT networks introduce additional complexities, such as managing high transaction volumes, optimizing limited computational and storage resources, and handling constantly changing network topologies.\par
Our enhanced mining process, therefore, is designed to address these specific challenges with four interconnected stages: transaction selection, miner assignment, transaction and block verification, and block encoding for storage efficiency. Each stage contributes distinct, targeted functionalities that improve scalability, resource optimization, and security within the IoT environment.\par
The process begins with selective transaction handling, where the BS prioritizes transactions based on factors like urgency, transaction fees, and age. This approach ensures efficient processing of high-priority transactions despite limited resources, as detailed in Section IV-A. Next, the BS assigns these prioritized transactions to a subset of reliable miners, carefully distributing the workload to prevent resource overload and foster scalability, as described in Section IV-B.\par
Following assignment, the miners verify transactions in a decentralized manner, ensuring that only valid transactions are added to the blockchain. This stage, crucial for maintaining blockchain integrity, also includes consensus formation and state updates, detailed further in Sections III-D and IV.\par
Finally, to address IoT storage constraints, we employ rateless Raptor codes to encode blocks, enabling miners to store blockchain data efficiently. This encoding approach allows each miner to retain only partial data fragments, which can later be reconstructed if needed, facilitating participation from storage-limited IoT devices as this strategy explored in Sections III-B and III-D.\par
By integrating these stages, our mining process extends beyond block generation to offer a robust framework optimized for the dynamic and resource-constrained nature of IoT networks. This structured approach ensures scalability, adaptability, and secure data handling within an IoT blockchain ecosystem.\par

\subsection{Proposed Mining Procedure}
In our proposed scheme, the BS is assumed to be a trusted entity that is always available. The BS collects transactions, performs transaction assignment, and selects miners. This entity is critical for coordinating the workflow but does not participate in mining or consensus itself.\par

The blockchain operates as a permissioned blockchain where all participating miners know one another. In this setting, each miner is aware of the identities of other miners in the network. Moreover, miners also know the set of assigned miners for each transaction batch. This transparency ensures that, as described in Section IV.B, every miner can collect and verify votes from other miners, facilitating a reliable and verifiable consensus process.\par

Our proposed mining scheme consists of four stages, as described in the following.\par
\textbf{Stage 1: Transaction Assignment:} At the beginning of epoch $t$ the BS collects $n$ transactions from different devices to form the matrix $T(t)=[L(t) \quad S(t) \quad R(t)]$. Next, the BS performs transaction selection and miner assignment, as follows.
\begin{enumerate}[label=(\roman*)]
\item Transaction selection: By solving an optimization problem, the BS selects $K(t)$ of the $n$ transactions that are more likely to be appended to the blockchain. By doing so, we try to minimize the amount of futile work in the network and speed up the verification process. Therefore, this will increase the throughput and lead to a shorter confirmation time for transactions, which is one of the most desired features in IoT blockchains.
\item Miner assignment: Based on the current network conditions, the BS will determine the number of the miners, $M(t)$, each selected transaction must be assigned to, such that the network can reach a consensus about the validity of each transaction with high probability. Then the BS will assign each selected transaction to $M(t)$ randomly selected miners.
\end{enumerate}
As the result of the above two steps, the BS finds the set of selected transactions $\mathcal{T}_{s}(t)=\{x_{{1}}, \ldots, x_{K(t)}\}$ corresponding to $K(t)$ rows of matrix $T(t)$, with the reduced matrix of the transactions $\hat{T}(t)=[\hat{L}(t) \quad \hat{S}(t) \quad \hat{R}(t)]$. Moreover, for each selected transaction $x_{i}, i=1, \ldots, K(t)$, a subset ${\cal M}_i$ of miners with $\left|\mathcal{M}_{i}\right|=M(t)$ is assigned for validating and further processing. Let $q_{j}(t)$ be the number of transactions assigned to miner $j$. Then on average each miner is assigned $q(t)=\frac{1}{N(t)}\sum_{j=1}^{N(t)}q_{j}(t)=K(t)M(t)/N(t)$ transactions to process. The details of the above two steps in Stage 1 are described in Section IV.\par
\textbf{Stage 2: Transactions Verification:} The BS sends to each miner the matrix of the assigned transactions $\hat{T}_{j}(t)=[\hat{L}_{j}(t) \mkern4mu \hat{S}_{j}(t) \mkern4mu \hat{R}_{j}(t)]$ consisting of $q_{j}(t)$ rows (transactions) from the reduced matrix of the transactions $\hat{T}(t)$. Using the notations in Section III-B, to check the validity of the assigned transactions, miner $j$ needs some blocks $\hat{\mathbf{B}}_{j}(\ell)$ from some of the past block groups $\ell$. Denote the set of all such required blocks $\hat{\mathbf{B}}_{j}(\ell)$ for verification at epoch $t$ by $\hat{\mathbf{B}}_{j}^{t-1}$. Recall that each miner stores a fraction of the coded data and uses it for verifying the assigned transactions. If some required data in $\hat{\mathbf{B}}_{j}^{t-1}$ are not locally stored by miner $j$, it contacts a subset of the miners in the network and decodes the intermediate blocks corresponding to the missing data in $\hat{\mathbf{B}}_{j}^{t-1}$. If such a repair process fails to recover $\hat{\mathbf{B}}_{j}^{t-1}$, miner $j$ then decodes all entire block groups that are associated with $\hat{\mathbf{B}}_{j}^{t-1}$.\par
Then having the required data $\hat{\mathbf{B}}_{j}^{t-1}$, miner $j$ computes $V_{j}^{t}(\hat{L}_{j}(t), \hat{S}_{j}(t),\hat{\mathbf{B}}_{j}^{t-1})$ to find the vector $v_{j}^{t} \in \{0,1\}^{q_{j}(t)}$ which is the verification results of miner $j$. Be noted that here the length of vector $v_{j}^{t}$ is $q_{j}(t)$, so each transaction is not verified by all miners. Then each miner sends the vector $v_{j}^{t}$ to other miners and the BS. Upon receiving the validation vectors from all miners, based on the miner assignment $\mathcal{M}(t)=\{\mathcal{M}_{1}, \ldots, \mathcal{M}_{K(t)}\}$ for the $K(t)$ selected transactions, each miner checks if for each transaction $x_{i}$, there are enough miners in $\mathcal{M}_{i}$ in favor of its validity, i.e., the majority of the miners in $\mathcal{M}_{i}$ accept $x_{i}$ as a valid transaction. In case of majority acceptance, $x_{i}$ is validated. As a result of the mentioned procedure, miners in the network can reach a consensus about the set of valid transactions $\hat{L}^{*}(t)$.\par
\emph{Remark 1}: Denote $d_{i}$ as the depth of the oldest block that is required for processing transaction $x_{i}$. That is if $x_{i}$ is generated at epoch $t$, then the oldest block required for processing $x_{i}$ is generated at epoch $t-d_{i}$. In the Transaction Selection step (see Section IV-A), the BS limits the depth of selected transactions, such that $d_{i}<D$. For example, for a batch of incoming transactions at epoch $t$, where $\ell W<t\leq (\ell+1)W$, if $D=t-\ell W$, then the BS selects transactions whose validations require blocks generated during the time interval $[\ell W,t]$, which are not encoded yet and are locally stored. All other transactions will be backlogged. On the other hand, if $D=t-(\ell-1)W$, then the required old blocks are generated during $[(\ell-1)W,t)$, which involves the most recently encoded blocks. And so on. Hence the BS can adjust $D$ based on the number of backlogged transactions and the new incoming transactions in each epoch so that the raw data required for validation can be successfully obtained through repairing, keeping the decoding complexity low.\par
\emph{Remark 2}: Sharding where a set of miners are partitioned into disjoint groups, and each group verifies a set of disjoint transactions, is a particular case of the mentioned scheme. In our scheme, the sets of miners for different transactions could be disjoint or have some members in common.\par
\emph{Remark 3}: The RNM algorithm employed here is the same as that in ~\cite{R5000} with a difference in usage. There the algorithm is used by newly joined miners only to store the ledger by recovering the previous blocks. On the other hand, in our scheme, in addition to the mentioned purpose, it is used to recover the blocks if they are required for transaction verification.\par
\emph{Remark 4}: In Stage 2, instead of broadcasting raw vote vectors \(v_j^t\), miners attach cryptographic hash commitments (e.g., using SHA-256) to their votes. This approach binds each miner to a unique vote value, so any attempt at equivocation, such as sending different votes to the BS versus other miners, would lead to mismatched hash values detectable by the network. Consequently, these cryptographic commitments ensure consistency and integrity of votes, preventing malicious discrepancies and preserving the reliability of the consensus process.\par
\textbf{Stage 3: Block Verification:}
At this step, each miner $j$ uses the set of valid transactions $\hat{L}^{*}(t)$ and forms the set $\hat{L}_{j}^{*}(t)$, which is a subset of the confirmed transactions from $\hat{L}^{*}(t)$, where the transactions that are not in $\hat{T}_{j}(t)$ are excluded. Denoting $\hat{R}_{j}^{*}(t)$ as the corresponding functions that are required to be called from the set $\hat{L}_{j}^{*}(t)$, miner $j$ finds the updated state of $\hat{S}_{j}(t)$ for its assigned contracts by computing ${F_{j}^{t}}(\hat{R}_{j}^{*}(t),\hat{S}_{j}(t))=\hat{S}_{j}(t)$, where ${F_{j}^{t}}$ is the function calling operator for miner $j$. The updated state will be broadcasted to the network and the BS. Similar to the previous stage, if the majority of the miners in $\mathcal{M}_{i}$ broadcasted the same updated state for transaction $x_{i}$, that state becomes the updated state for $x_{i}$. This way, the updated state $\hat{S}^{*}(t)$ for all confirmed transactions at Stage 2 are determined. Each miner appends the block $\hat{B}(t)=[\hat{L}^{*}(t) \quad \hat{S}^{*}(t) \quad \hat{R}^{*}(t)]$ to its local storage until the moment of the group block encoding that is explained in the next stage. Also, the BS updates the status of the network and reliability of each miner by comparing the received results of computations with the collective decisions on the validity of transactions which will be detailed in Section IV-B.\par
\textbf{Stage 4: Block Encoding:} Recall that in our scheme the blocks are encoded in groups of size $W$. Suppose that $\hat{B}((\ell-1)W+1)$ is the first block in the $\ell$-th group of blocks. Stages 1-3 repeat for all epochs $t=(\ell-1)W+1$ through $t=\ell W$. After the generation of the block $\hat{B}(\ell W)$, it is the time for encoding the $\ell$-th block group, which are blocks $\{\hat{B}((\ell-1)W+1), \hat{B}((\ell-1)W+2), \ldots, \hat{B}(\ell W)\}$. The group size $W$ is determined by the BS and based on the number of active nodes in the network. First, as the pre-coding step, miners in the network encode blocks $\{\hat{\mathbf{b}}_{\ell}(1), \ldots, \hat{\mathbf{b}}_{\ell}(W)\}$ to form the intermediate codewords $\{\hat{\mathbf{d}}_{\ell}(1), \ldots, \hat{\mathbf{d}}_{\ell}(\overline{W})\}$. Then following the procedure in Section III-B each miner $j$ generates and stores one coded block $\mathbf{c}_{\ell}(j)$. Afterward, miners erase all the original blocks $\{\mathbf{b}_{\ell}(1), \ldots, \mathbf{b}_{\ell}(W)\}$. But, even though it is not necessary, depending on the storage capability of each miner, they could keep the intermediate blocks $\{\hat{\mathbf{d}}_{\ell}(1), \ldots, \hat{\mathbf{d}}_{\ell}(\overline{W})\}$ for some time, so that when these blocks are required for transactions verification, they are directly available without repairing the corresponding coded block. Hence it is an instance of trading storage for computing.\par
It is worth mentioning that in any stage during the generation of the blocks $\left\{\hat{B}(i)\right\}_{i=(\ell-1)W+1}^{\ell W}$ if miners need any data from blocks that are older than $\hat{B}((\ell-1)W+1)$, they need to contact other miners and perform the repair process or decoding to obtain the data. However, if the required data is in the blocks $\left\{\hat{B}(i)\right\}_{i=(\ell-1)W+1}^{i=\ell W}$ because the encoding of the $\ell$-th group of blocks has yet to be started, that data are stored locally by each miner as raw data.\par
Also, note that due to the use of rateless codes, given the set of data blocks in one group as the source symbols, one can generate potentially unlimited number of encoded symbols. Specifically, each newly joined miner $j^\prime$ can generate and store $\mathbf{c}_{\ell}(j^\prime)$ by first generating the degree number $L_{j^\prime}$ from degree distribution $\Omega(L)$ and then selecting $L_{j^\prime}$ intermediate blocks from the set $\{\hat{\mathbf{d}}_{\ell}(1), \ldots, \hat{\mathbf{d}}_{\ell}(\overline{W})\}$ to form its coded block. Therefore, our scheme adapts to dynamic networks, and miners can join and leave the network without having an adverse effect on the encoding and decoding procedures. Such a feature is especially well suited for IoT networks which are typically highly dynamic. However, other coding schemes, such as the LCC-based coded blockchain, are designed for a network with a fixed number of nodes.\par
Furthermore, the reason for choosing the raptor codes as the coding scheme for our blockchain, in addition to being a linear time encoding and decoding scheme, is that it has a very low decoding failure probability, i.e., the probability that a miner fails to decode a group of blocks is very low, which makes it a suitable candidate for IoT blockchains with strict security and timing requirements.\par
The proposed mining procedure for the coded IoT blockchain is detailed in Algorithm~\ref{alg:coded_mining}, outlining the sequential stages of transaction assignment, verification, consensus, and block encoding.\par

\begin{algorithm}[ht]
\caption{Coded IoT Blockchain Mining Procedure}
\label{alg:coded_mining}
\SetAlgoLined

\textbf{Stage 1: Transaction Assignment} \\
Collect $n$ transactions $T(t) = [L(t), S(t), R(t)]$ \\
Assign transactions and miners (Refer to Algorithm~\ref{alg:transaction_selection}) \\

\textbf{Stage 2: Transaction Verification} \\
\ForEach{miner $j$ assigned transactions $\hat{T}_j(t)$}{
    Retrieve required prior blocks $\hat{\mathbf{B}}_j^{t-1}$ \\
    Verify transactions: $v_j^t = V_j^t(\hat{L}_j(t), \hat{S}_j(t), \hat{\mathbf{B}}_j^{t-1})$ \\
    Broadcast verification result $v_j^t$ to other miners and the Base Station (BS)
}
\textbf{Consensus:} Finalize valid transactions $\hat{L}^*(t)$ \\

\textbf{Stage 3: Block Verification} \\
\ForEach{miner $j$}{
    Execute valid contracts: $\hat{S}_j(t) = F_j^t(\hat{R}_j^*(t), \hat{S}_j(t))$ \\
    Broadcast updated state $\hat{S}_j(t)$
}
\textbf{Consensus:} Update global state $\hat{S}^*(t)$ \\
Form block: $\hat{B}(t) = [\hat{L}^*(t), \hat{S}^*(t), \hat{R}^*(t)]$ \\

\textbf{Stage 4: Block Encoding} \\
\If{$t = \ell W$}{
    Encode blocks $\{\hat{B}((\ell-1)W+1), \ldots, \hat{B}(\ell W)\}$ \\
    Store encoded block $\mathbf{c}_\ell(j)$ for each miner \\
    Discard raw blocks $\{\hat{B}((\ell-1)W+1), \ldots, \hat{B}(\ell W)\}$
}

\end{algorithm}

\section{Transaction Assignment} In this section, we describe the transaction selection and the miner assignment steps in the first stage of our proposed mining procedure for coded IoT blockchain.\par

\subsection{Transaction Selection} Our goal is to select a subset of $n$ transactions to maximize the total reward subject to constraints on computing and storage resource as well as the depth of required blocks for processing transactions. Three factors determine the reward of a transaction:
\begin{enumerate}
\item \emph{Vitality}: It is an indicator of the level of urgency for the transaction. A higher value means the transactions should be processed faster.
\item \emph{Transaction fee}: With the same vitality level, it is preferred to first work on transactions with higher fees to maximize the profit of the miners.
\item \emph{Age}: It is the time between the submission of the transaction and sending the transaction to the miners for processing. If a transaction does not have a high priority or a competitive transaction fee, it should not be prevented from being sent for processing because many backlogs can result in network congestion.
\end{enumerate}
We use Richards' curve~\cite{R33} to model the reward of each transaction based on the above three factors. Denote $v_{j}$, $a_{j}$, and $f_{j}$ as the vitality, age of the transactions in number in epochs, and transaction fee, respectively, of the $j$-th transaction, $j=1, \ldots, n$. Denoting $\textbf{v}=(v_{1}, \ldots,v_{n})$, and $\textbf{a}=(a_{1}, \ldots,a_{n})$, the reward of the $j$-th transaction is given by:
\begin{equation}\label{eq:reward_function}
r_{j}(\textbf{v},\textbf{a},f_{j})=\frac{1}{\left(1+\tilde{v}_{j}e^{-f_{j}}\right)^{1/\tilde{a}_{j}}},
\end{equation}
where
\begin{equation}\label{eq:reward_function_2}
\tilde{v}_{j}=\frac{e^{\frac{-1}{\lVert \textbf{v} \rVert} v_{j}}}{\sum_{k=1}^{n} e^{\frac{-1}{\lVert \textbf{v} \rVert} v_{k}}}, \qquad \tilde{a}_{j}=\frac{e^{\frac{1}{\lVert \textbf{a} \rVert} a_{j}}}{\sum_{k=1}^{n} e^{\frac{1}{\lVert \textbf{a} \rVert} a_{k}}}.
\end{equation}
Fig. 2 shows the efficiency of the proposed model for transaction prioritization. For $n=5$ with $\textbf{v}=(1, 10, 6, 8, 2, 4)$ and $\textbf{a}=(16, 1, 4, 32, 2, 8)$, we plot $r_{j}$ as a function of $f_{j}$ for $j=1, 2, 3$. From the figure we can see in the high fee region, e.g., $f_{j}>10$, the transaction with high vitality ($j=2$) has higher priority and approaches the maximum faster. On the other hand, in the lower fee region, its priority becomes lower than the transaction with older age ($j=1$).\par
\begin{figure}[!htb]
\centering
\includegraphics[width=3.75in]{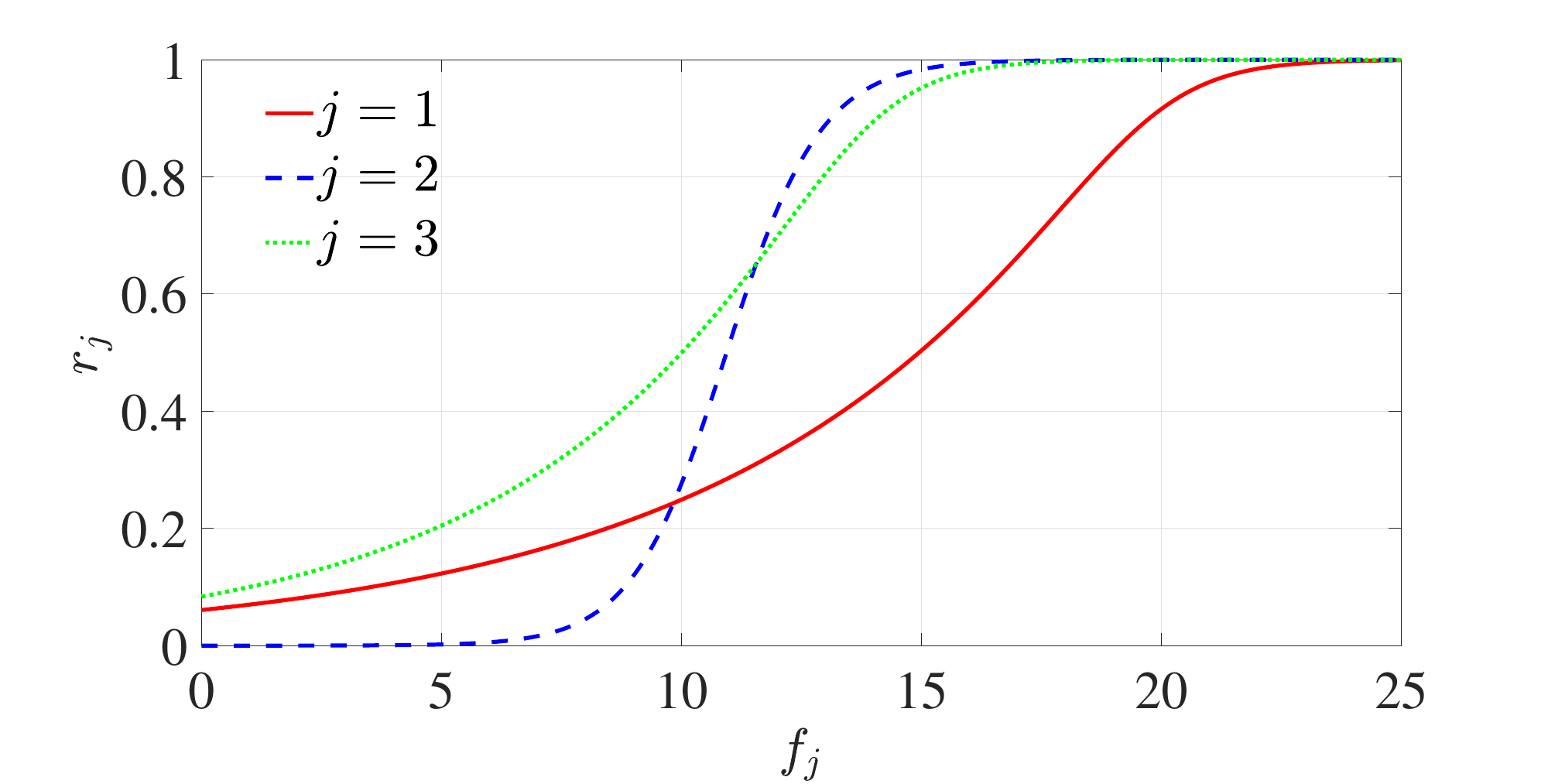}
\caption{Plot of the reward function $r_{j}(\textbf{v},\textbf{a},f_{j})$.}
\label{uncalib}
\end{figure}
For each transaction $j$, define $\xi_{j}$ as the amount of computation resource required for processing it, modeled as a Gamma random variable with the shape parameter $\alpha_{j}$ and scale parameter $\beta$; define $\eta_{j}$ as its size, modeled as a Gaussian random variable with mean $\mu_{j}$ and variance $\sigma^{2}$; and define $d_{j}$ as the maximum depth, which is the position of the oldest block required for processing transaction $j$, modeled as a Poisson random variable with mean $\lambda_{j}$. Furthermore, define the binary variable $x_{j}=1$ if the $j$-th transaction is selected for processing and $x_{j}=0$ otherwise. Then the problem of selecting a subset of transactions can be formulated as the following optimization problem:
\begin{subequations}\label{eq:optimization}
\begin{align}
\underset{~x_{j} \in\{0,1\},\,j=1, \ldots,n}{\operatorname{\max}} & \quad \sum_{j=1}^{n} r_{j} x_{j}, \\
s.t.:~~~~~~~~&\hspace{-1cm}\mathbb{P}\left(\sum_{j=1}^{n} \xi_{j} x_{j}\leq C\right)=P\left(\sum_{j=1}^{n} \alpha_{j} x_{j}, \frac{C}{\theta}\right)\geq q_{1}, \label{eq:comp_constraint} \\
&\hspace{-1cm}\mathbb{P}\left(\sum_{j=1}^{n} \eta_{j} x_{j}\leq S\right)=\Phi\left(\frac{S-\sum_{j=1}^{n} \tau_{j} x_{j}}{\sqrt{\omega \sum_{j=1}^{n} \tau_{j} x_{j}}}\right)\geq q_{2}, \label{eq:size_constraint} \\
&\hspace{-2.5cm}\mathbb{P}\left(\text{max}(d_{1}x_{1}, \ldots, d_{n}x_{n})\leq D\right)=\prod_{j=1}^{n}Q(D+1,\lambda_{j}x_{j})\geq q_{3}. \label{eq:depth_constraint}
\end{align}
\end{subequations}
where $P$, $\Phi$, and $Q$ are the lower regularized gamma function, the cdf of standard normal and the upper regularized gamma function, respectively. Formulation (\ref{eq:optimization}) maximizes the total reward of the selected transactions subject to three probabilistic constraints: (\ref{eq:comp_constraint}) puts a constraint on the computational cost for the selected transactions; (\ref{eq:size_constraint}) limits the aggregated size of the selected transactions to the size of one block; and (\ref{eq:depth_constraint}) restricts the depth of the required blocks for processing the selected transactions.\par
Since in an IoT blockchain, the number of transactions is typically much higher than that in conventional blockchains like cryptocurrency blockchains, the traditional mining process in which each miner processes all incoming transactions becomes inefficient. Here we cluster the incoming transactions for faster and more efficient processing leading to higher throughput. First, based on the transactions' type and calling functions, they would consume a wide range of computing resources; hence, they need to be judiciously chosen to meet the computing capacity limit. Second, due to the diversity of the transactions and contracts in the IoT blockchain, the size of each transaction could be in a broader range than conventional blockchains. Hence if a transaction is not likely to be accommodated in a block along with other submitted transactions, it will not be selected. Third, by restricting the depth of selected transactions to only a few previous blocks, in case miners could not repair the required blocks and had to decode a group, we limit the total decoding complexity.\par
Alternatively, when $\xi_{j}$, $\eta_{j}$, and $d_{j}$ are known to the BS, we can also consider the following deterministic optimization formulation:
\begin{subequations}\label{eq:optimization_deterministic}
\begin{align}
\underset{~x_{j} \in\{0,1\},\,j=1, \ldots,n}{\operatorname{\max}} & \quad \sum_{j=1}^{n} r_{j} x_{j},
\\
s.t.:~~ & \sum_{j=1}^{n} \xi_{j} x_{j}\leq C,\\
& \sum_{j=1}^{n} \eta_{j} x_{j}\leq S,\\
& \text{max}(d_{1}x_{1}, \ldots, d_{n}x_{n})\leq D.
\end{align}
\end{subequations}
Problems (\ref{eq:optimization}) and (\ref{eq:optimization_deterministic}) are stochastic and deterministic forms of the well-known knapsack problem ~\cite{R34,R35,R36} in combinatorial optimization, respectively, which are NP-complete~\cite{R37}. We relax them by changing the constraints $x_{j} \in\{0,1\}$ to $x_{j} \in[0,1]$. It is shown in~\cite{R9000} that the stochastic knapsack problem in (\ref{eq:optimization}) is a convex optimization problem in its relaxed form. Moreover, the relaxed version of the deterministic problem (\ref{eq:optimization_deterministic}) is a linear program. Finally, the solution to the relaxed problem can be mapped to a binary solution by using the random rounding method~\cite{R38}.\par
Note that the stochastic formulation (\ref{eq:optimization}) and the deterministic one in (\ref{eq:optimization_deterministic}) have their own advantages and drawbacks. Formulation (\ref{eq:optimization}) is more demanding computationally, but the BS does not need to know the exact values of different transaction attributes, i.e., size, amount of required resource for computation, and depth of the oldest required block for verification. On the other hand, though formulation (\ref{eq:optimization_deterministic}) is less demanding computationally, the BS must know the attribute values of each transaction which may not be available for some IoT blockchains.
\subsection{Miner Assignment} In our scheme, the BS keeps monitoring the reliability of each miner. Specifically, denote $p_{j}(t)$ as the probability that miner $j$ is reliable, estimated by the BS at epoch $t$. Recall that $v^{*}(t)\in \{ 0,1\}^{K(t)}$ is the binary vector whose $i$-th element is $1$ if the majority of miners validate the $i$-th selected transaction by the BS, and $0$ otherwise. Also, $v_{j}(t)\in \{0,1\}^{q_{j}(t)}$ contains votes of miner $j$ at epoch $t$ about the validity of its set of assigned transactions. Denote $v_{j}^{*}(t)\in \{ 0,1\}^{q_{j}(t)}$ as the subvector of $v^{*}(t)$ that includes just the corresponding elements of $v_{j}(t)$. Now assume that at time epoch $t-1$, the reliability of miner $j$ is estimated as $p_{j}(t-1)$. Then after receiving all votes from all miners in the network and knowing the collective decision about the validity of each transaction, the BS can update the reliability of miner $j$ as follows:
\begin{equation}\label{eq:average_reliability}
p_{j}(t) = (1-\beta) p_{j}(t-1) + \beta\frac{l_{j}(t)}{q_{j}(t)}
\end{equation}
where $0<\beta<1$ is a forgetting factor to count for time-evolution, and $l_{j}(t)$ correct votes of miner $j$ at epoch $t$, i.e., the number of zeros in $v_{j}^{*}(t)\oplus v_{j}(t)$. Now, suppose that at the beginning of epoch $t$ the BS wants to assign each transaction from $\hat{T}(t)$ to $M(t)$ miners in the network in a way that after each miner collects the votes of other miners, the network is able to reach a consensus about the validity of transactions, based on the majority rule, with a probability of at least $1-\epsilon$ for $0<\epsilon<1$. Denote $Z_{j}(t)= 1$ if miner $j$'s vote is reliable and 0 otherwise. Now denote $Z(t)= Z_{1}(t) + \ldots+ Z_{M(t)}(t)$. The network will make a correct decision if $Z(t)>\frac{M(t)}{2}$. Approximating the miners’ votes as i.i.d. Bernoulli random variables with parameter $P(t)=\left(\prod_{j=1}^{N(t)}p_{j}(t)\right)^{1/N(t)}$, and using the Chernoff bound~\cite{R251} we have:
\begin{equation}
\operatorname{Pr}\left[Z(t)>\frac{M(t)}{2}\right] \geq 1-e^{-\frac{1}{2P(t)} M(t)\left(P(t)-\frac{1}{2}\right)^{2}} \geq 1-\epsilon\,,
\end{equation}
which results in:
\begin{equation}\label{eq:M_t}
M(t)=\left\lceil\frac{8(\ln 1/\epsilon) P(t)}{(1-2P(t))^{2}}\right\rceil
\end{equation}
as the minimum number of miners required to be assigned to each transaction. Therefore, the BS computes $P(t)$ and $M(t)$ at the beginning of each epoch $t$ and randomly assigns each of the $K(t)$ selected transactions from the reduced matrix of transactions $\hat{T}(t)$ to $M(t)$ miners.\par
The process of transaction selection and miner assignment is described in Algorithm~\ref{alg:transaction_selection}, where transactions are evaluated based on a reward function, optimized under resource constraints, and assigned to miners.
\begin{algorithm}[ht]
\caption{Transaction Selection and Miner Assignment}
\label{alg:transaction_selection}
\SetAlgoLined

\KwIn{Transactions $\{1,2,\ldots,n\}$ with vitality $v_j$, age $a_j$, fee $f_j$, resource requirements $\xi_j$, sizes $\eta_j$, depths $d_j$}

\textbf{Compute Rewards:} \\
\ForEach{transaction $j$}{
    Compute reward $r_j$ using equation~\eqref{eq:reward_function}
}

\textbf{Formulate and Solve Optimization Problem:} \\
Maximize total reward $\sum_{j=1}^n r_j x_j$ subject to constraints in equations~\eqref{eq:comp_constraint}, \eqref{eq:size_constraint}, and \eqref{eq:depth_constraint} \\
Solve for $x_j^*$ \\
\textbf{Select} transactions where $x_j^* = 1$ to form the set $\mathcal{T}_s(t)$ \\
\textbf{Compute} average miner reliability $P(t)$ using equation~\eqref{eq:average_reliability} \\
\textbf{Determine} number of miners per transaction $M(t)$ using equation~\eqref{eq:M_t} \\

\ForEach{transaction $x_i \in \mathcal{T}_s(t)$}{
    Randomly assign transaction $x_i$ to $M(t)$ miners to form $\mathcal{M}_i$
}

\end{algorithm}

\section{Comparison with Existing Coded Blockchains} In this section, we will compare our proposed scheme with two other benchmark papers on coded blockchains, which are Polyshard~\cite{R14} and LCB~\cite{R4} in terms of several performance metrics, including storage reduction, decentralization, throughput, and security. Although both papers use the Lagrange polynomial to code the blockchain data, Polyshard uses a sharded blockchain, while LCB is based on a non-sharded approach. The reason for choosing these two papers is that to the best of our knowledge, at the time of writing this paper, these two works are the only benchmarks on coded blockchains that fit IoT applications.\par
\subsection{Simulation Setup} We analyze a dynamic network where miners join and leave at the transition between epochs. The number of miners joining or leaving follows Poisson distributions with means $\lambda_e$ and $\lambda_l$, respectively. In the pre-code stage, we employ the Reed-Solomon code from the Matlab 2021a Communication Toolbox. The size of encoded groups in our scheme, $W_{\ell}$, is dynamic and varies for each group based on the number of miners in the network at encoding time.\par
The $\ell$-th group of encoded blocks consists of $\{\hat{B}(W_{(\ell-1)}+1), \ldots, \hat{B}(W_{\ell})\}$, and its size is determined to maintain a pre-code rate of $W_{\ell}/\overline{W} = 0.8$. To calculate $W_{\ell}$, we use the failure probability function from ~\cite{R5000}, which estimates the probability of failing to decode at least one input symbol for a rateless code given its generator matrix $\mathbf{G}$ and degree distribution $\Omega(L)$. By determining the overhead required to ensure reliable decoding with high probability and using the current number of miners $N(t)$ (equal to the number of output coded symbols), we compute $W_{\ell}$ such that the failure probability remains below a predefined threshold. The modified robust soliton distribution ~\cite{R7000} is used as the degree distribution given by: 
\begin{equation}
\Omega(L)= \begin{cases}0, & \text { for } L=1 \\ \mu(L)+\frac{\mu(1)}{\overline{W}-1}, & \text { for } L=2, \ldots, \overline{W}\end{cases}
\end{equation}
where
\begin{equation}\label{cccuuu}
\mu(i)=\frac{\rho(i)+\tau(i)}{\beta}, \quad \text { for } \quad 1 \leq i \leq \overline{W}
\end{equation}
and $\beta=\sum_{i=1}^{\overline{W}}(\rho(i)+\tau(i))$. In (\ref{cccuuu}), $\rho(i)$ and $\tau(i)$ are given by:
\begin{equation}
\begin{gathered}
\rho(i)=\left\{\begin{array}{cl}
1 / \overline{W}, & \text { for } i=1, \\
1 /(i(i-1)), & \text { for } 2 \leq i \leq \overline{W},
\end{array}\right. \\
\tau(i)=\left\{\begin{array}{cl}
\mathcal{S} / i \overline{W}, & \text { for } 1 \leq i \leq \frac{\overline{W}}{\mathcal{S}}-1, \\
\mathcal{S} \ln (\mathcal{S} / \delta) / \overline{W}, & \text { for } i=\frac{\overline{W}}{\mathcal{S}}, \\
0, & \text { otherwise. }
\end{array}\right.
\end{gathered}
\end{equation}
where  $\delta \in[0,1]$ and $c>0$ are constant, and $\mathcal{S}=c \cdot \sqrt{\overline{W}} \cdot \ln \left(\frac{\overline{W}}{\delta}\right)$ is the average number of degree-1 code symbols. We set $c=0.15$ and $\delta=0.5$. As seen in Fig. 3, this degree distribution favors lower degrees more than higher degrees, which helps reduce the decoding complexity.\par
\begin{figure}[!htb]
\centering
\includegraphics[width=3.75in]{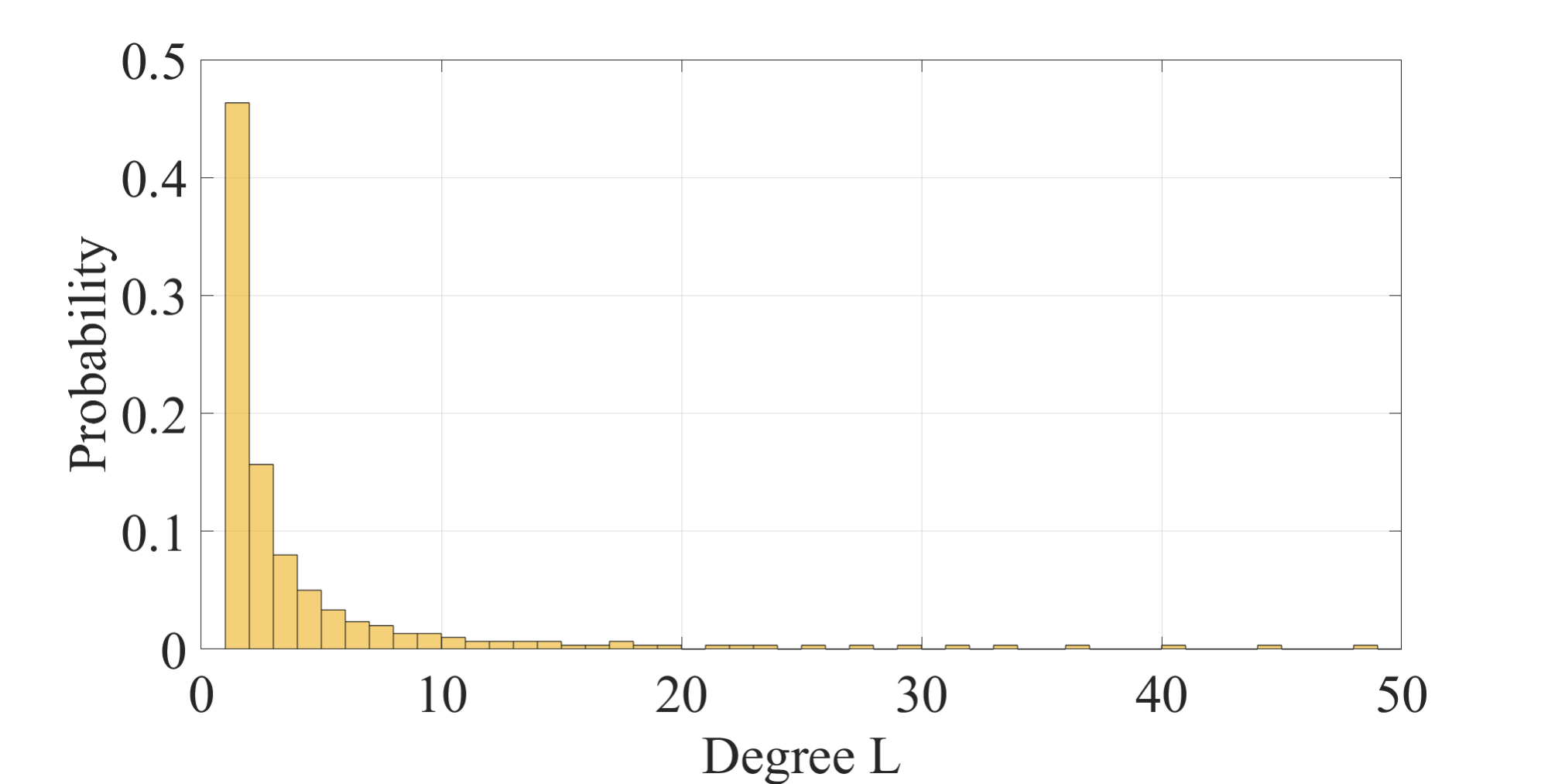}
\caption{The modified robust soliton distribution used in the LT code ($\overline{W}=1000, \delta=0.15, c=0.5)$.}
\label{degree}
\end{figure}
Our simulation has three types of miners: dishonest, honest, and straggler. A dishonest miner always votes incorrectly, while an honest miner always successfully delivers correct votes to the network. Also, a straggler is a miner that, with probabilities of $1/3$ and $2/3$, does not send any votes and sends correct votes to other miners, respectively. To have a fair comparison with LCB and Polyshard, adopting the parameters in ~\cite{R4}, we assume that in all three schemes, up to $40\%$ of miners in the network can be a straggler.\par  

We distinguished straggler nodes from malicious ones to capture realistic network behavior and apply more tailored strategies, ensuring that benign delays are not mistaken for adversarial actions. Malicious nodes intentionally disrupt consensus by sending false information, whereas benign straggler nodes may delay or fail to send responses due to network latency, resource constraints, or temporary failures without malicious intent. This distinction allows our model to differentiate between non-malicious performance issues and genuine attacks, targeting security measures appropriately.\par
Building on the threat model of stragglers and malicious nodes described above, we assume that \emph{up to} $40\%$ of nodes in the network are configured as potential stragglers to represent worst-case performance scenarios. However, statistically, not all of these nodes experience delays simultaneously, resulting in an effective number of stragglers that remains well below $40\%$. Combined with scenarios involving up to $30\%$ malicious nodes, the aggregate proportion of non-responsive or adversarial nodes is highly likely to stay below $50\%$, preserving a majority of honest and responsive nodes. This ensures that our majority voting mechanism and consensus process remain reliable and secure, even under adverse conditions.\par

The simulations are run on a Java-based simulator, JABS~\cite{R6000}, a modular blockchain network simulator with the capability of handling around $10000$ nodes. For our optimization in Stage 1, we use GAMS~\cite{R6001} and solve the problem of transaction selection on its Java API, which has built-in support for functions in our problem, including regularized gamma functions and standard normal cdf. We assume that there is enough bandwidth for sending and receiving the data required for consensus, and the latency is negligible compared to the block size. Also, to choose the system parameters we used the Ethereum live data set~\cite{R6002} and ~\cite{R6003, R6004}, and set our simulation parameters as follows.\par
At the beginning of each epoch, the BS collects $n=500$ transactions to solve the problem in (\ref{eq:optimization}) to select a subset of them for further processing by miners. This batch of $n=500$ transactions are either all newly generated transactions or a combination of new and backlogged transactions. Denote the number of backlogged transactions at the beginning of epoch $t$ by $\Upsilon(t)$. The number of the newly created transactions at the beginning of epoch $t$ is then $500-\Upsilon(t)$. For each transaction $j$ we generate six numbers representing its fee $f_{j}$, age $a_{j}$, vitality $v_{j}$, amount of computation resource required for processing $\xi_{j}$, size $\eta_{j}$, and maximum depth $d_{j}$ following the distributions described below. In case a transaction is backlogged in an epoch, it will preserve these attributes in future epochs until it is selected for processing by the BS.\par
We assume that a transaction's fee and age are related to its vitality as follows: $f_{j}{\sim} {\rm Exp}(20/v_{j})$, truncated between $0$ to $100$; $a_{j}{\sim} {\rm Exp}(6/v_{j})$, truncated between $1$ to $32$. For equations (\ref{eq:reward_function}) and (\ref{eq:reward_function_2}), the vitality  $v_{j}$ of transaction $j$ is uniformly distributed in $\{1,2, \ldots, 10\}$ with 10 being the highest vitality. Also, $\xi_{j}{\sim} {\rm Gamma}(1.5,42000)$, $\eta_{j}{\sim}\mathcal{N}(3000, 1000)$. Denote the last confirmed block before the start of epoch $t$ by $\mathcal{W}$. Then at the beginning of epoch $t$, $500-\Upsilon(t)$ new transactions must be generated. We divide the $500-\Upsilon(t)$ transactions into five groups of equal size with different maximum depths. For transaction $j$ in group $\kappa$, $d_{j}{\sim}{\rm Poisson}(\mathcal{W}(0.95-0.23\kappa))$ for $\kappa=0, 1, \ldots, 4$. Moreover, in problem (\ref{eq:optimization}) the values of $C=6.7\times 10^{6}$, $S=1.2 \times 10^{6}$ are fixed for all the epochs. However, the value of $D$ changes in each epoch according to the number of backlogged transactions. That is, if the majority of the batch of $500$ transactions for selection by the BS in problem (\ref{eq:optimization}) are from the backlogged transactions, we continue to increase $D$ and set it to its next value from the set of values $\{t-W_{\ell}, t-W_{(\ell-1)}, \ldots\}$ until the majority of the batch of $500$ transactions are new transactions.\par
Furthermore, in equation (\ref{eq:average_reliability}), the initial value of $p_{j}$ for miner $j$ when it first joins the system is uniform between $0.5$ and $1$, $\beta=0.1$, and $\epsilon=0.01$ in equation (\ref{eq:M_t}).\par
Finally, for the proposed scheme, the simulated networks in Sections V-B and V-C are dynamic, and Section V-C considers a non-dynamic network.\par
\subsection{Storage}
In IoT blockchains, miners often overlap with users submitting transactions. Due to the diverse capabilities of IoT users, some may struggle with storage as the blockchain grows. General solutions to blockchain storage issues typically involve pruning~\cite{R39} (ignoring old block data) or using side-chains~\cite{R40}, both of which introduce drawbacks~\cite{R41}. Here, we compare our scheme to Polyshard and LCB regarding storage usage fraction $R_{s}$, defined as the storage space needed per node in a coded blockchain relative to an uncoded conventional blockchain.\par
In LCB, all miners store identical coded blocks, so coding does not reduce storage. Hence, the storage usage fraction is $R_{s} = 1$ for LCB. In sharded blockchains, the number of shards depends on the network size and the fraction of malicious miners. In Polyshard, storage constraints are also influenced by the verification function's polynomial degree. For Polyshard, $R_{s} = \frac{1}{\lfloor \frac{(1-2\mu)N(t)-1}{U} \rfloor}$~\cite{R14}, where $\mu$ is the fraction of dishonest miners and $U$ is the maximum degree of the verification function.
The storage usage fraction of our scheme is given by:
\begin{equation}
    R_{s}=\frac{Q-\sum_{\ell \in \mathcal{W}} W_{\ell}+|\mathcal{W}|}{Q},
\end{equation}
where $Q$ is the last confirmed block, $\mathcal{W}$ is the set of encoded blocks, and $W_{\ell}$ is the number of consecutive blocks in the $\ell$-th group of encoded blocks.\par
Fig. 4 shows the evolution of the storage usage fraction $R_s$ over time in our scheme compared to Polyshard and LCB. We simulated a dynamic blockchain where the numbers of nodes joining and leaving each epoch follow Poisson distributions with $\lambda_l = 4$ and $\lambda_e = 10$, respectively. Starting with 1000 miners, we calculated $R_s$ across 250 epochs.\par
Initially, $R_s = 1$ because blocks are not encoded until the first encoding instant. After encoding starts, miners need significantly less storage to store coded blocks, leading to a steep decline in $R_s$. Fluctuations in $R_s$ occur because miners temporarily store all blocks in a group until the BS encodes them. Once encoded, miners erase these blocks and keep only a coded block, reducing $R_s$. This fluctuation repeats with every group.\par
For Polyshard, $R_s$ for the balance-checking function~\cite{R14} is shown with $U=1$, and for $\mu=30\%$ and $\mu=40\%$ (fractions of dishonest miners). In Polyshard, $R_s$ decreases as the fraction of dishonest nodes increases, while in our scheme, $R_s$ remains unaffected by the number of dishonest miners. Thus, with few dishonest miners, Polyshard and our scheme show comparable $R_s$. However, as the fraction of dishonest miners grows, our scheme achieves greater storage savings.\par
Additionally, Polyshard's storage usage increases with higher degrees of the verification polynomial (larger $U$), and the comparison here is based on Polyshard's optimal case, $U=1$.\par
\begin{figure}[!htb]
\centering
\includegraphics[width=3.75in]{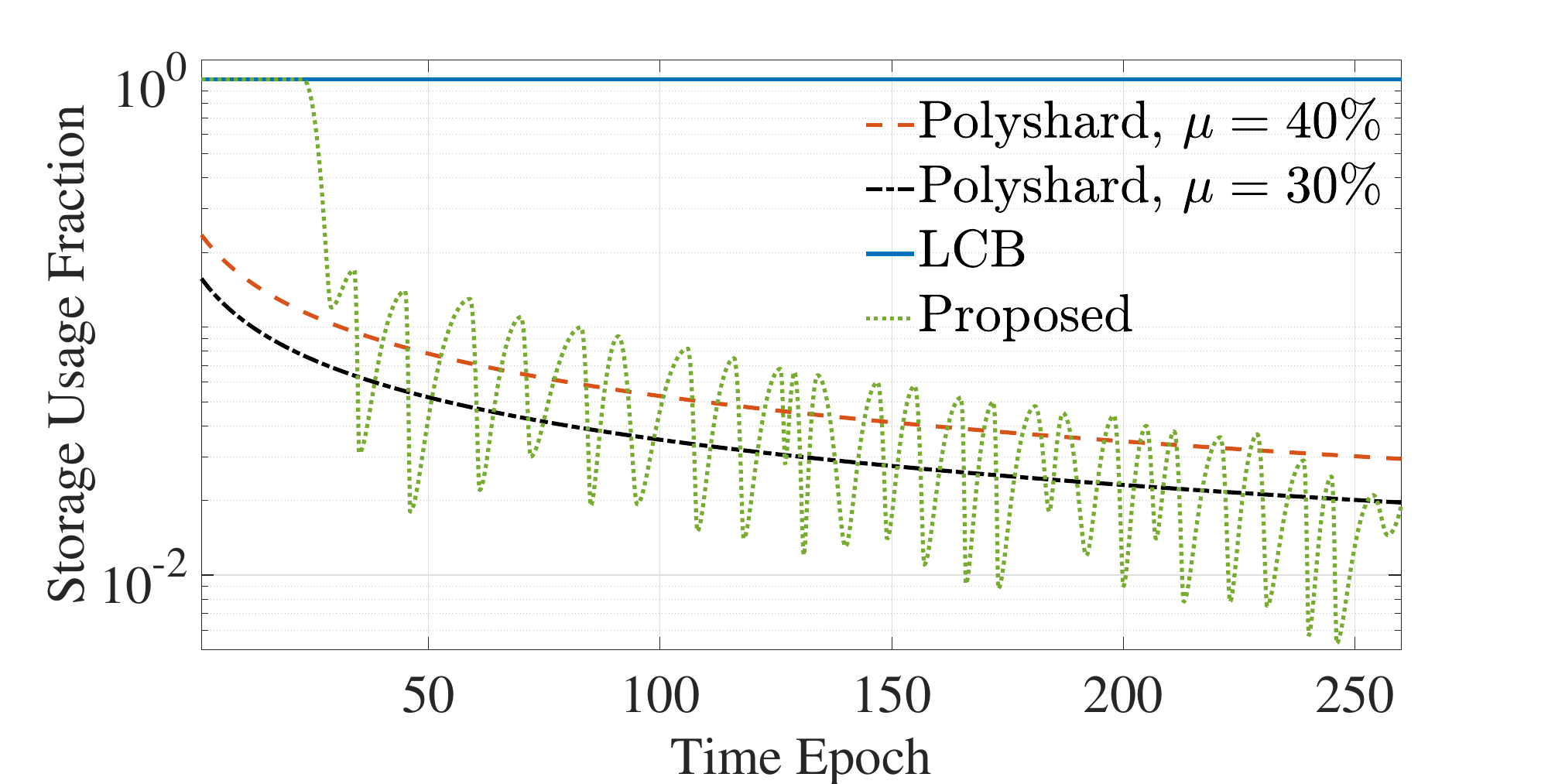}
\caption{Storage usage fraction comparison with $N =1000$, $\lambda_{l}=4$, and $\lambda_{e}=10$.}
\label{thglrd31}
\end{figure}
\subsection{Decentralization}
We differentiate between decentralized and distributed systems. Decentralized systems involve decision-making across multiple nodes, with each node determining its behavior and collectively influencing the system, ensuring no single node holds absolute decision-making power. Distributed systems, however, refer to processing spread across nodes, but decision-making may still be centralized or decentralized.\par
In both Polyshard and LCB, computation is distributed but not truly decentralized. For Polyshard, block generation is unclear, and computations are only distributed during verification. If a conventional consensus algorithm (e.g., PoW, PoS, PBFT) is used for block generation, Polyshard becomes centralized due to the inherent centralization issues of these algorithms ~\cite{R42}.\par
In LCB, a well-equipped node can undertake multiple mining tasks, increasing its contribution to block generation or verification and pushing the network toward centralization. We compare decentralization in LCB with our scheme, while for Polyshard, analysis is limited to descriptive observations due to its ambiguous block generation process.\par
Denote $\phi_{j}(t)$ as the number of blocks generated by miner $j$ at epoch $t$. We adopt the following two metrics to measure decentralization ~\cite{R43}. The Gini coefficient at epoch $t$ is defined as:\par
\begin{equation}\label{gg}
I(t)=\frac{\sum_{i=1}^{N(t)} \sum_{j=1}^{N(t)}\left|\phi_{i}(t)-\phi_{j}(t)\right|}{2 N(t)\sum_{j=1}^{N(t)} \phi_{j}(t)}.
\end{equation}
To provide intuitive understanding, the Gini coefficient is a statistical measure originally used to assess income inequality. In our context, it evaluates how evenly block generation is distributed among miners. A \emph{lower Gini coefficient} signifies that blocks are generated more uniformly across miners, reflecting higher decentralization since no single miner or small group dominates the process. Conversely, a \emph{Gini coefficient close to 1} indicates that block generation is concentrated among a few miners, pointing to centralization.\par
On the other hand, the entropy at epoch $t$ is defined as:\par
\begin{equation}\label{entee}
E(t)=-\sum_{j=1}^{N(t)} \frac{\phi_{j}(t)}{\sum_{j=1}^{N(t)} \phi_{j}(t)} \log _{2} \frac{\phi_{j}(t)}{\sum_{j=1}^{N(t)} \phi_{j}(t)}.
\end{equation}
To further elucidate, \emph{entropy} is a concept from information theory that measures the randomness or unpredictability within a system. In our scenario, higher entropy indicates a more random and evenly spread distribution of mining power among miners. This randomness ensures that block generation is less predictable and more uniformly distributed, signifying greater decentralization. \emph{Higher entropy} thus reflects that no single miner holds a significant advantage, whereas \emph{lower entropy} suggests predictability and potential centralization.\par
We simulated the LCB and our proposed scheme for 500-time epochs with $N=500$ miners in a fast dynamic network with $\lambda_{l}=10$ and $\lambda_{e}=20$. For our scheme, denote by $\mathcal{K}_{v}(t)$ the set of transactions out of $K(t)$ selected transaction by the BS that is verified by the network and appended to the chain as one block which corresponds to the $1$'s in $v^{*}(t)$. Therefore, for any $x_{i} \in \mathcal{K}_{v}(t)$, we give participation credit to miner $j \in \mathcal{M}_{i}$, if miner $j$ voted in favor of validity of $x_{i}$. Finally, the sum of all credits that miner $j$ receives from all $x_{i} \in \mathcal{K}_{v}(t)$ form $\phi_{j}(t)$. For the LCB scheme, assume that the time that miner $j$ takes to finish the assigned tasks at epoch $t$ follows an exponential distribution with mean $h_{j}(t)$. Denoting the number of total tasks at epoch $t$ by $H(t)$, then the number of assigned tasks to miner $j$ for epoch $t$ is $a_{j}(t)=\frac{1/h_{j}(t)H(t)}{\sum_{j=1}^{N(t)}1/h_{j}(t)}$. Hence miners with smaller delays are assigned more mining tasks. Then following ~\cite{R4}, we set a deadline to finish the assigned tasks, and all returned tasks after the deadline will be discarded. We set $\phi_{j}(t)$ as the percentage of successfully finished assigned tasks by miner $j$ from all finished tasks by all miners, indicating its participation in decision-making and decentralization. Adopting the parameters in the simulation section of~\cite{R4}, $H(t)=10+3.2N(t)$, the deadline for returning the assigned task is $5$, $h_{j}(t){\sim}\mathcal{N}(4.5,1.66)$ and $h_{j}(t){\sim}\mathcal{N}(2.5,5/8)$ for straggler and non-straggler miners, respectively. Also, in rare cases, if $h_{j}(t)<0$, we assume that the assigned task for miner $j$ is returned on time. Fig. 5 (a) shows the histograms for the Gini coefficients for the two schemes. It is seen that our scheme has a much smaller Gini coefficient than LCB. In fact, the Gini coefficient of our scheme is close to zero, indicating excellent decentralization. Moreover, in Fig. 5 (b), the histograms of the entropies of the two schemes are shown. It is seen that our scheme has a larger mean and larger variance than LCB, indicating a better decentralization.\par
\begin{figure}
\centering
\begin{minipage}{0.48\textwidth}
\includegraphics[width=3.75in]{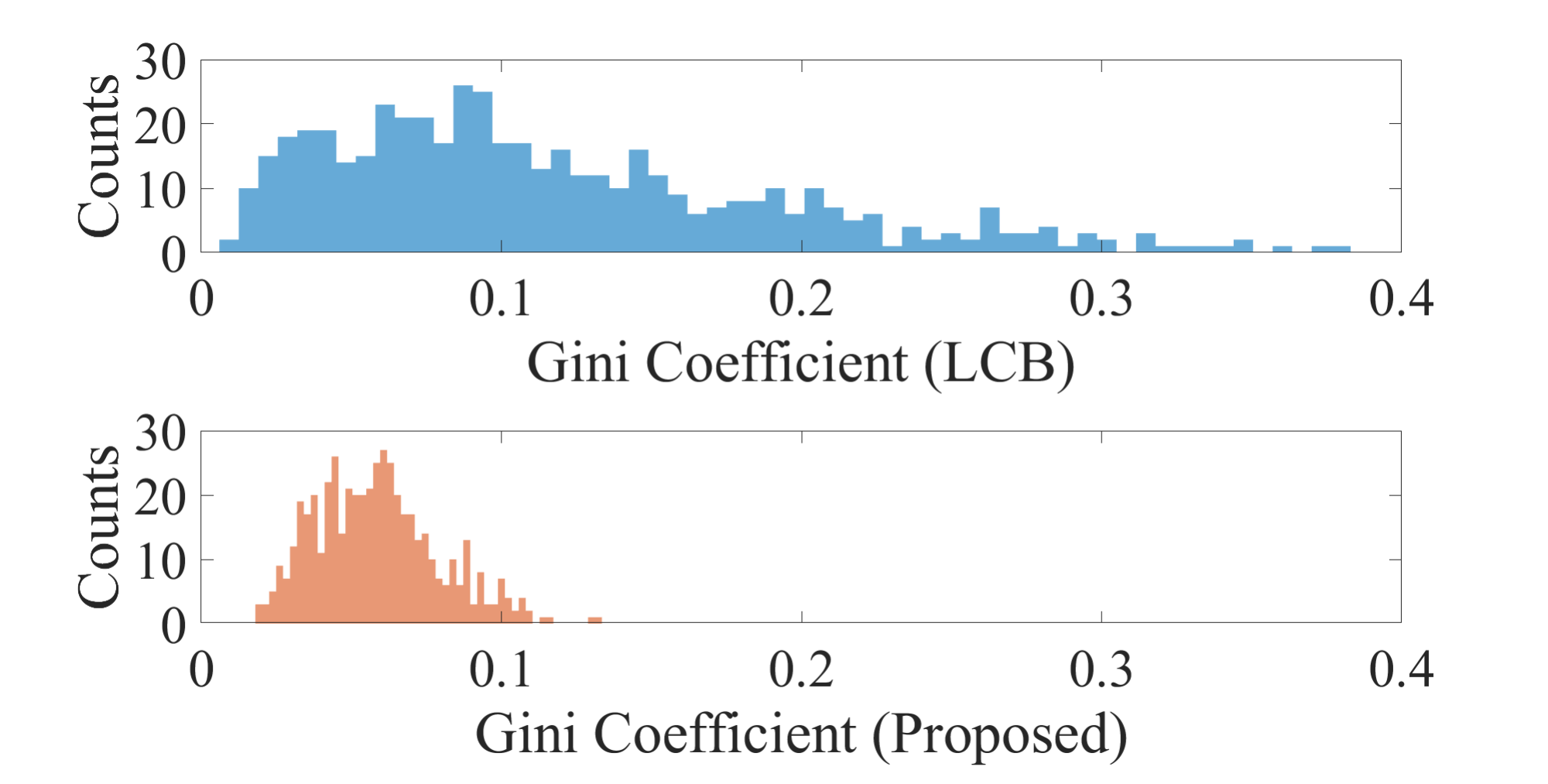}
\caption*{(a)}
\end{minipage} \hfill
\begin{minipage}{0.48\textwidth}
\includegraphics[width=3.75in]{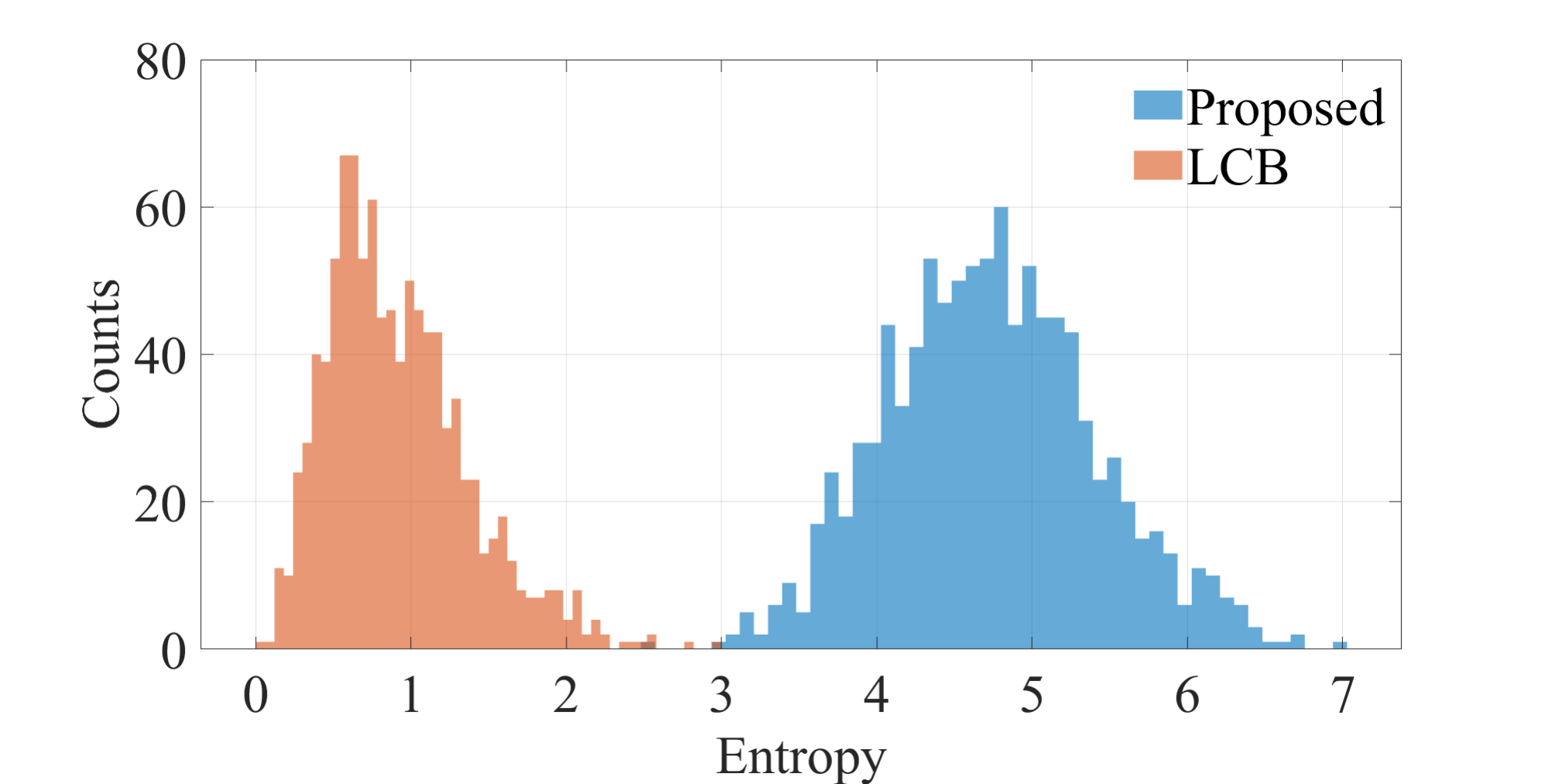}
\caption*{(b)}
\end{minipage}
\caption{The comparison of decentralization between LCB and the proposed scheme with $N=500$, $\lambda_{l}=10$, $\lambda_{e}=20$ and for 500-time epochs, a) Gini coefficient, and b) entropy.}
\end{figure}
\subsection{Throughput} Throughput is defined as the average number of transactions correctly confirmed per epoch. In uncoded blockchains, throughput primarily depends on the runtime of consensus algorithms. For example, in PoW blockchains, the time required to find the target nonce for solving the hash puzzle determines transaction confirmation time. In coded blockchains, however, coding and decoding complexities significantly influence confirmation time and throughput. Next, we explain how the throughput of the three schemes was obtained through simulations.\par
For all three schemes, the network has a dishonest miner percentage $\mu=30\%$, with up to $40\%$ straggler miners as described in Section V-A. Simulations involve a ledger size of $10000$ epochs.\par 
In Polyshard, the verification function has degree $U=1$ (balance-checking). The number of shards is $S=\lfloor (1-2\mu) N-1 \rfloor\in[15,165]$, based on the miner count $N$. Each shard proposes a block with $\frac{500}{S}$ transactions. Following the Polyshard scheme, miners generate and broadcast intermediate results. These results address stragglers (erasure) and dishonest miners (error) via Reed-Solomon decoding (dimension $S$, length $N$). A block is valid if the network collectively confirms its validity for each shard $i\in[1, S]$. Let $\varrho(N)$ represent the total confirmed transactions across all valid blocks for $N$ miners. Polyshard's throughput is calculated as $\frac{\bar\varrho(N)}{500}$, where $\bar\varrho(N)$ is averaged over $1000$ simulation rounds.\par 
In LCB, block generators and validators are equal in number. Following the setup in~\cite{R4}, the block generation/validation polynomial degree is $2$. The number of block generation and verification tasks is $N+5$ (based on equations (5) and (6) in~\cite{R4}), meeting the security and resiliency threshold of LCC. Tasks are assigned as per Section V-C, using the same straggler distributions. Each miner decodes coded blocks and verifier results to confirm blocks. For block $B(t)$, verifier $j$'s vote $\Psi_{j}(t)\in\{0,1\}$ determines confirmation via the consensus rule in~\cite{R4}: $B(t)$ is valid if decoded successfully and $\sum_{j=1}^{N}\frac{a_{j}(t)}{H(t)}\Psi_{j}(t)\geq0.5$, where $a_{j}(t)$ and $H(t)$ are defined in Section V-C. LCB throughput is the average confirmed transactions per block over $1000$ rounds, divided by $500$.\par 
In our scheme, throughput is based on correctly confirmed transactions $\mathcal{K}_{v}(t)$ from a batch of $n=500$ transactions, simulated with and without transaction selection. Without transaction selection, the optimization in Section IV-A is skipped, and transactions are distributed using equation (\ref{eq:M_t}). With selection, the BS solves problem (\ref{eq:optimization}) to choose a subset of transactions. Similar to Polyshard and LCB, throughput is averaged over $1000$ rounds of simulations.\par
In Fig. 6, we show the throughputs for the three schemes as a function of the number of miners $N$. The throughput of the proposed scheme is higher than Polyshard and LCB. A couple of reasons led our scheme to have higher throughput than the other schemes. One is that we run the blockchain transaction-based, so the validity status of one transaction does not impact other transactions. While in the two other schemes having just one invalid transaction in one block causes the whole block to be invalid, thus making all effort and time spent in vain. Another reason is that in our scheme, we use raptor codes with a linear time encoding and decoding complexity, while the other two schemes employ LCC, which has a higher computational complexity than raptor codes. For instance, the complexity of decoding a length-$N$ Reed-Solomon code at each node to find the collective decision regarding the validity of $S$ proposed is $\mathcal{O}(N\log^{2}N\log\log N)$~\cite{R14}. However, the complexity of finding $W_{\ell}$ raw blocks in one group, which are coded by raptor code, is $\mathcal{O}(\frac{N}{(1+\varepsilon)}\log(1/\varepsilon))$, where $\varepsilon$ is the threshold for probability of failure~\cite{R4000}. Lower complexity helps the computation for each miner take less time, leading to a higher throughput.\par
Moreover, it seems that the transaction selection in stage 1 of the proposed scheme can significantly increase the throughput since results in faster and less futile work. Also, though when the number of miners is small, e.g., $N=50$, the proposed scheme without transactions selection has a lower throughput than LCB and Polyshard, for networks with a large number of miners, e.g., $N>400$, it outperforms LCB and Polyshard. The reason is that the slop of the throughput curve for the proposed scheme without transactions selection is higher than that of LCB and Polyshard.\par
Finally, it should be noted that the increase in throughput as the number of miners grows reflects the inherent scalability of our sharding-inspired design, which assigns subsets of transactions to different miners and processes them in parallel. As the network expands, more miners can operate concurrently on distinct transaction subsets, effectively reducing bottlenecks and speeding up transaction validation and block creation. This parallel transaction processing is analogous to sharded blockchain systems, where increasing shards or processing units leads to higher aggregate throughput. Our scheme further optimizes this process through coded computation, dynamic miner assignment, and prioritized transaction selection, which together mitigate coordination overhead and computational latency. Consequently, despite the intuition that a larger network might complicate coordination, the system leverages additional resources to enhance overall performance, resulting in the observed increase in throughput with a growing number of miners.

\begin{figure}[!htb]
\centering
\includegraphics[width=3.75in]{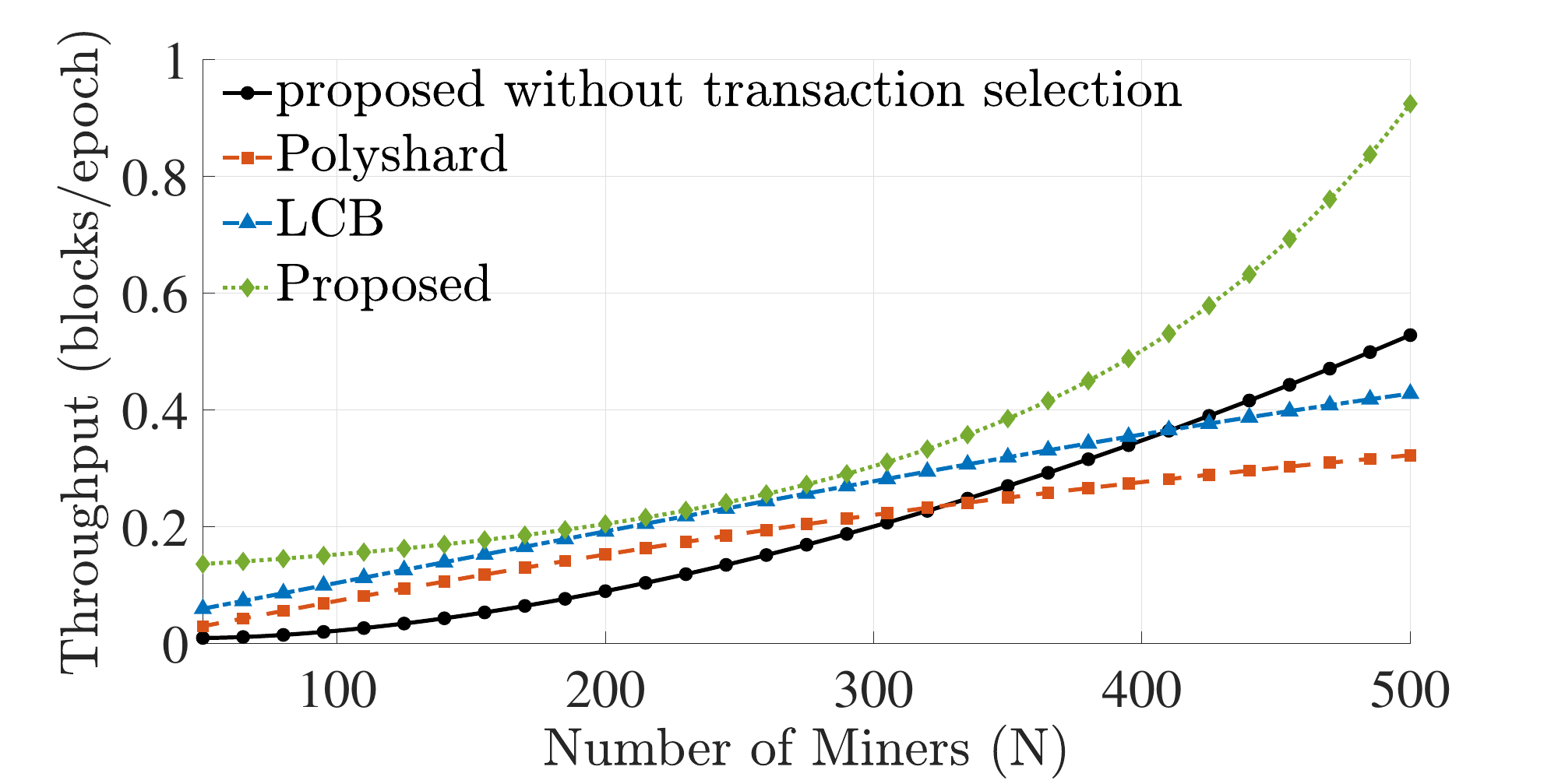}
\caption{Throughput versus the number of miners $N$ with the percentage of dishonest miners $\mu=30\%$.}
\label{time}
\end{figure}
Fig. 7 illustrates how the percentage of dishonest miners, $\mu$, impacts throughput in a non-dynamic network. We measured throughput for a fixed $N=500$ miners with $\mu$ ranging from $5\%$ to nearly $50\%$, each over $1000$ simulation rounds.\par
Our proposed scheme maintains steady throughput as $\mu$ increases from $5\%$ to $30\%$ and then declines more rapidly from $30\%$ to $50\%$, yet it remains significantly higher than the other two schemes. For LCB, we simulated two scenarios based on the distributions in Section V-C by sorting $h_{j}(t)$ for all miners. In the first scenario, $80\%$ of honest miners are in the first half of the list and $20\%$ in the second, meaning more powerful miners are mostly honest. In the second scenario, $20\%$ of honest miners are in the first half and $80\%$ in the second, making most powerful miners dishonest. This demonstrates the effect of computational power distribution in LCB. Results show that in the first scenario, LCB maintains similar throughput despite more malicious miners, while in the second scenario, its throughput drops below Polyshard when $\mu > 40\%$.
\begin{figure}[!htb]
\centering
\includegraphics[width=3.75in]{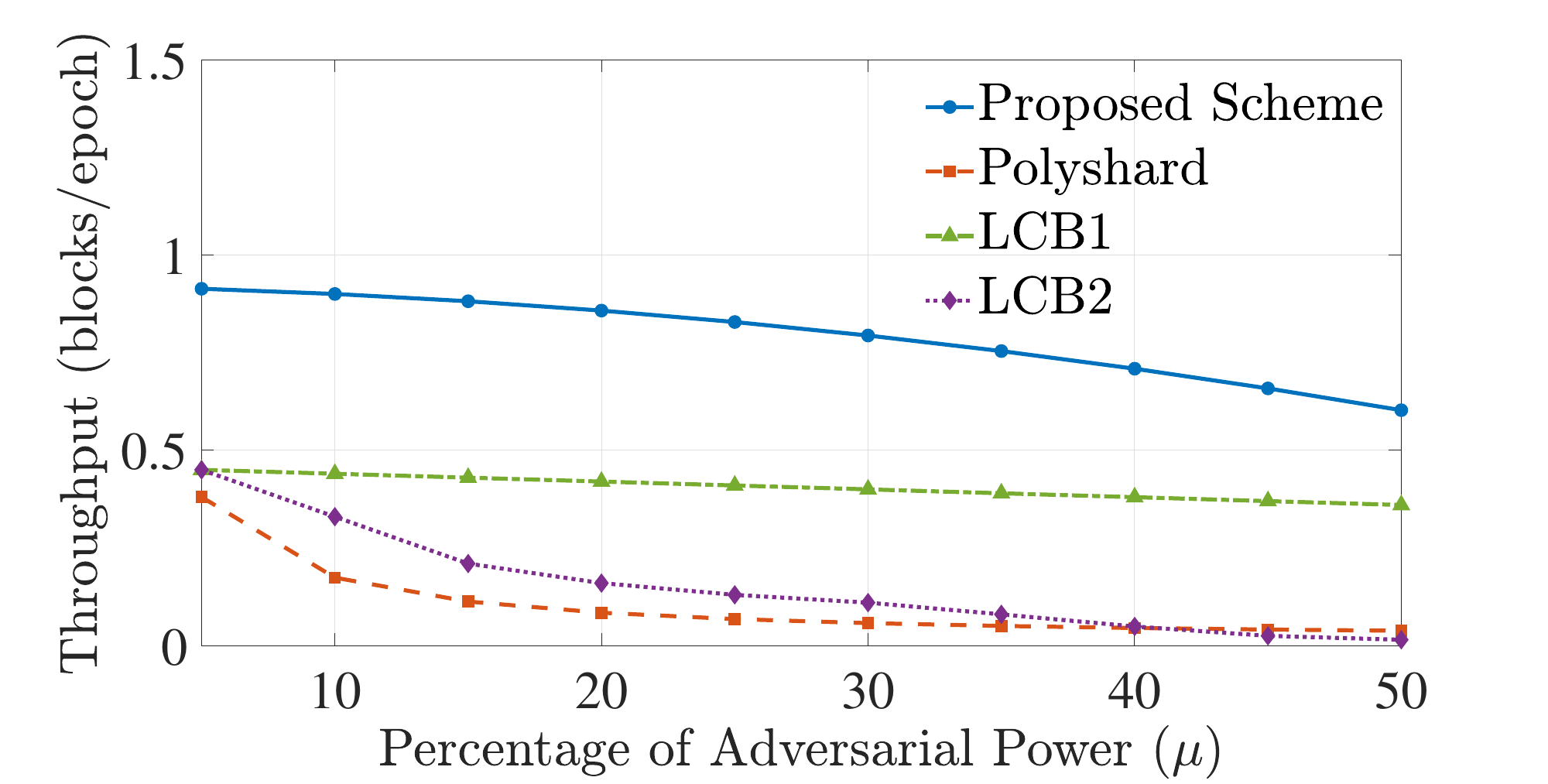}
\caption{Throughput versus the percentage of dishonest miners $\mu$ with $N=500$ miners.}
\label{thglrd31}
\end{figure}
\subsection{Security} 
Blockchain security has many aspects but can be broadly categorized as malicious and non-malicious behaviors. Malicious behavior involves miners deliberately disrupting the network or appending incorrect blocks. Non-malicious behavior leads to similar outcomes but occurs unintentionally, due to failures or miner straggler effects.\par
\subsubsection{Discrepancy Attack} The discrepancy attack~\cite{R28} occurs during block generation. Polyshard assumes all nodes receive the same blocks at the start of each epoch, with blocks proposed for each shard by nodes within that shard. Sharding, which divides the blockchain into parallel sub-chains by splitting miners into smaller groups, makes the network vulnerable to attacks since smaller groups are easier to control. An adaptive adversary can target a shard and generate multiple conflicting blocks, causing network discrepancies. Detecting such discrepancies~\cite{R28} involves significant communication costs. Preventing this requires the network's miners to scale as $S^{\frac{U}{2}}$, where $S$ is the shard count and $U$ the polynomial verification degree—a challenge for real-world blockchains.\par
The LCB counters this attack using coding in block generation. All miners propose a single block per epoch, formed using the LCC. Thanks to the error resilience of LCC, the network can reliably identify the correct block despite discrepancies.\par
In our proposed scheme, discrepancy attacks can occur at two levels: transaction and block. At the transaction level, attackers send conflicting votes on a subset of transactions to different miners. At the block level, they send conflicting blocks.\par 
To address transaction-level attacks, our scheme selects the number of miners $M(t)$ to validate each transaction based on the miners' estimated reliability during Stage 1. This ensures that even with conflicting votes, the scheme can reach a correct collective decision.\par
For block-level attacks, all miners broadcast the confirmed block, and the block agreed upon by the majority is appended to the chain. A successful block-level attack would require attackers to control more than half of the network miners, which is unlikely in real-world blockchains.\par
\subsubsection{Straggler Effect} In both LCB and Polyshard, consensus involves sub-tasks such as computing intermediate results of Lagrangian polynomials at specific points. Polyshard assumes all miners return their results on time. While coding techniques can handle some erasures by treating straggler miners as such, an increase in stragglers reduces the network's ability to handle erroneous results from adversarial miners.\par
LCB addresses this by assigning multiple computation tasks to each miner. Miners with greater computing power handle more tasks, compensating for delays caused by stragglers who fail to deliver results on time.\par
Our scheme mitigates straggler effects by adjusting the number of miners assigned to validate each transaction based on Equation (\ref{eq:M_t}). It accounts for stragglers or other network conditions that delay miners by continuously updating their reliability. As the number of stragglers increases, the scheme assigns more miners to process each transaction, ensuring the network can still make accurate decisions.\par
\subsubsection{Eclipse and 51\% Attacks} An eclipse attack involves an attacker creating a controlled environment around one or more miners, manipulating them into wrongful actions. A $51\%$ attack occurs when over half of the miners are controlled by a single group, allowing them to dominate decision-making in the network.\par
Both Polyshard and LCB use LCC, but they differ in handling encoded outputs. Polyshard waits for all encoded results before decoding, while LCB allows each miner to decode using a subset of the fastest coded outputs. This approach makes LCB more vulnerable to eclipse and $51\%$ attacks, especially in IoT blockchains where traditional consensus methods like PoW are absent, making it easier and cheaper to produce erroneous results. By assigning multiple tasks to some miners, LCB risks malicious miners targeting subsets to send incorrect outputs. LCB assumes that faster miners are mostly honest, which may not hold in practice, allowing powerful malicious miners to corrupt the chain by taking on a significant number of tasks.\par
In contrast, Polyshard assumes secure block generation within shards and sufficiently large shard sizes to prevent $51\%$ attacks, making it resistant to these threats under such assumptions.\par
In our scheme, random transaction assignment at the BS stage prevents attackers from targeting specific miners, as they cannot predict which miners will validate a transaction. Additionally, decisions rely on majority rule at the block verification stage, making a $51\%$ attack successful only if adversaries control more than half of all miners, which is highly improbable.\par
\subsubsection{Attacks in Distributed Encoding} Both LCB and Polyshard use distributed encoding, where miners exchange data, apply LCC, and store the resulting coded ledger locally. Miners then process transactions on the coded data and use LCC for consensus. However, as shown in~\cite{R32}, if dishonest miners send conflicting data to honest miners in a distributed encoding network, honest miners can be confused about the final decision, potentially causing the blockchain to fork at each epoch. Preventing this requires every honest miner to contact all others and wait for their results, negating the performance advantages of waiting only for the fastest miners' outputs in LCB and Polyshard.\par
Our scheme, however, does not rely on distributed encoding as described in~\cite{R32}. Instead, computations are performed on raw data, and each miner encodes past block data independently. This approach makes our scheme resistant to the attack described in~\cite{R32}.\par
Finally, we summarize whether or not each coded scheme is resistant to various attacks in Table 1.\par
\begin{table}[htb]
\centering
\hspace*{-0.5cm}\begin{tabular}{|c"c|c|c|c|}
\hline
\textbf{Attack Type} & \textbf{Polyshard} & \textbf{LCB} & \textbf{Proposed} \\ \thickhline
\textbf{Discrepancy Attack} & No & Yes & Yes \\ \hline
\textbf{Straggler Effect} & No & Yes & Yes \\ \hline
\textbf{Eclipse and 51\% Attacks} & Yes & No & Yes \\ \hline
\textbf{Attacks in Distributed Encoding} & No & No & Yes \\\hline
\end{tabular}
\caption{Resistance to various attacks by different coded blockchain schemes.}
\end{table}
\subsubsection{Simulation Comparison}
We conducted two experiments to evaluate the impact of attacks on the three schemes. The first experiment examined the effect of a discrepancy attack by measuring throughput as a function of the number of miners, similar to Section V-D, but without straggler miners. All miners were either honest or dishonest, with $20\%$ ($\mu=20\%$) being dishonest and sending discrepant values to disrupt the network.\par 
We defined $\vartheta$ as the number of distinct messages a dishonest miner can send during the block generation stage~\cite{R28}. For Polyshard and LCB, $\vartheta$ represents the number of different coded blocks a dishonest miner can send. Unlike simulations where dishonest miners send just one invalid block ($\vartheta=1$), they sent multiple invalid blocks ($\vartheta>1$) to corrupt consensus and hinder block decoding.\par
Our scheme, however, differs fundamentally. Transactions are processed directly, with blocks created only after valid transactions are identified. Consequently, $\vartheta$ is irrelevant in our scheme. Votes on transaction validity are binary: dishonest miners can either incorrectly vote valid transactions as invalid or invalid ones as valid, but they cannot send multiple discrepant messages. This binary voting mechanism eliminates the ability to exploit $\vartheta$ for a discrepancy attack in our scheme.\par
We computed the throughput of LCB, Polyshard (for $\vartheta=\{2,3\}$), and the proposed scheme, with results shown in Fig. 8. Polyshard is vulnerable to attacks, as an increase in miners leads to more dishonest ones (fixed $\mu$), causing more discrepant values and complicating LCC decoding~\cite{R28}. This reduces successfully decoded blocks and throughput, worsening as $\vartheta$ rises. For networks with around $300$ miners, Polyshard’s throughput drops to near zero.\par
LCB performs better, leveraging LCC for both block generation and verification. While the attack slightly decreases throughput due to increased decoding complexity with more miners/messages, the decline is slower than in Polyshard, even as $\vartheta$ rises.\par
The proposed scheme is immune to discrepancy attacks by processing transactions on raw data, where dishonest miners can only send false votes for their assigned transactions. Unlike Polyshard and LCB, where invalid transactions in blocks affect others, the proposed scheme processes each transaction separately, maintaining throughput even as miners increase.\par
In the second experiment, with $N=500$ miners and varying straggler percentages, throughput was analyzed (Fig. 9). Polyshard showed the largest throughput decline, as it assumes timely shard proposals and task completion. LCB demonstrated stable throughput despite rising stragglers, effectively addressing the issue. The proposed scheme maintained consistent throughput due to its miner assignment method (Section IV-B).\par

\begin{figure}[!htb]
\centering
\includegraphics[width=3.75in]{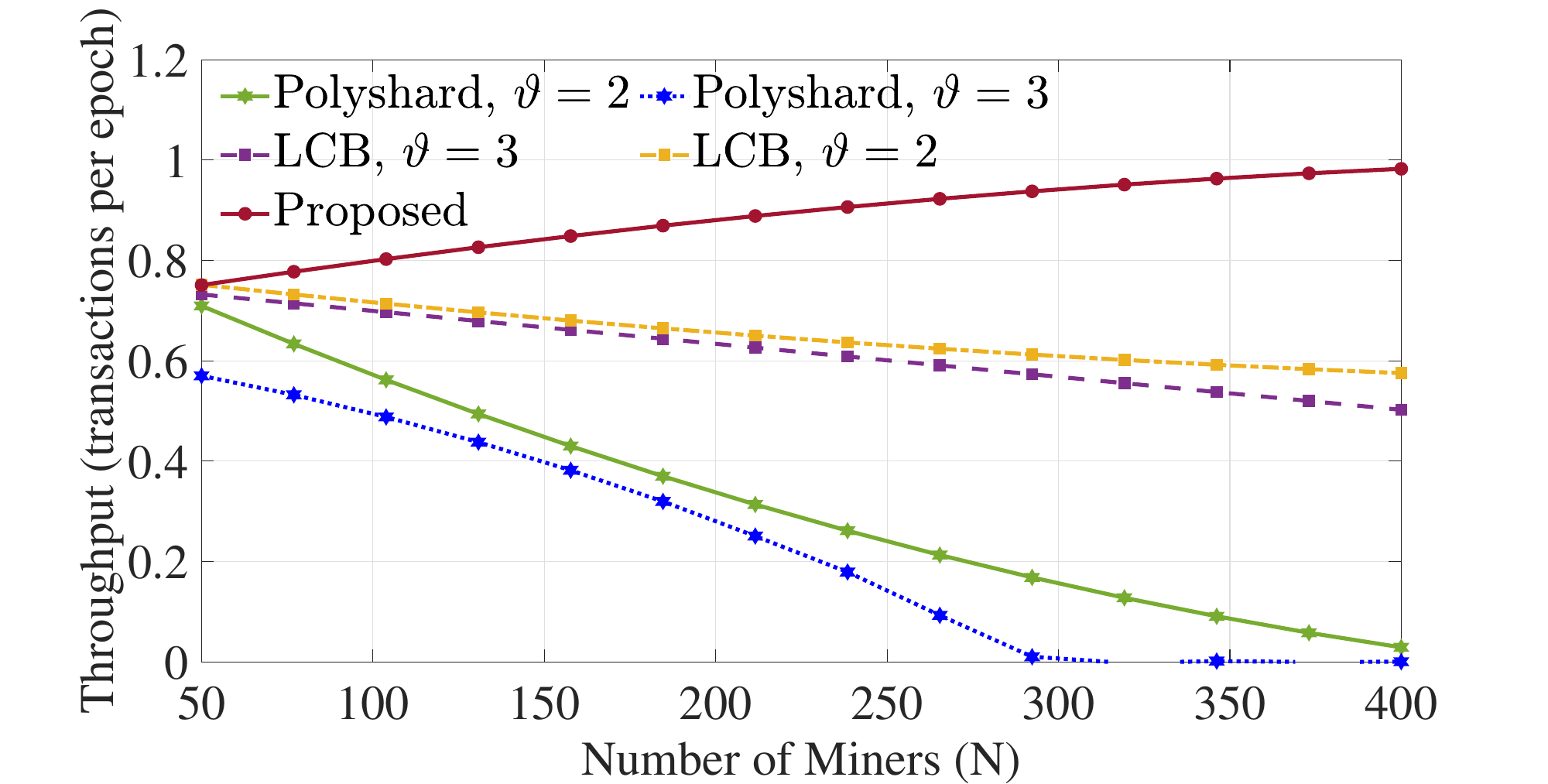}
\caption{The effect of discrepancy attack on the throughput based on the number of the miners ($N$) with $\mu=20\%$.}
\label{discrepancy}
\end{figure}

\begin{figure}[!htb]
\centering
\includegraphics[width=3.75in]{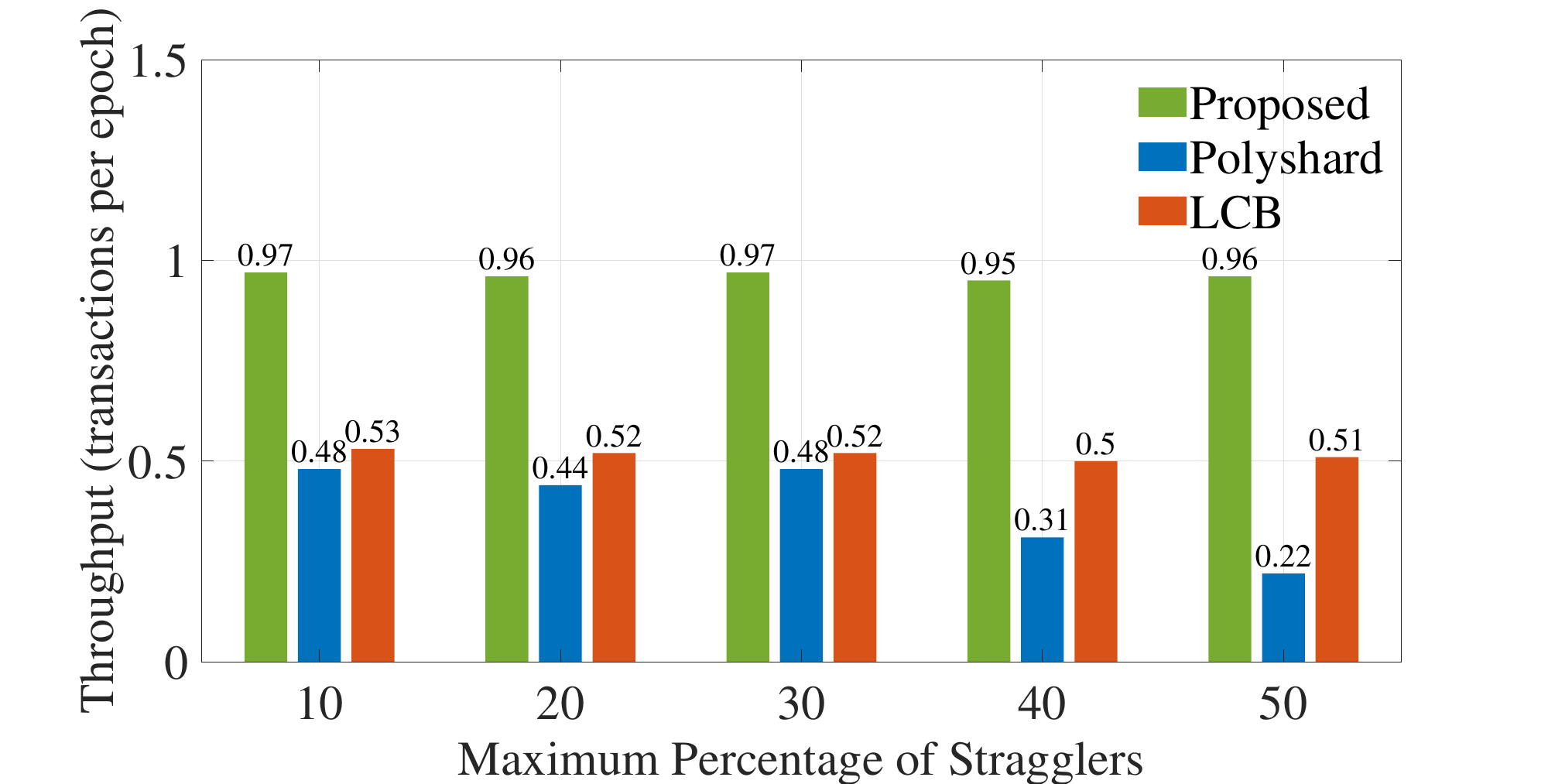}
\caption{The effect of stragglers on the throughput based on the maximum percentage of the straggler miners with $N=500$.}
\label{straggler}
\end{figure}

\section{Conclusions} We have proposed a new coded IoT blockchain to simultaneously improve the throughput, security, decentralization, and storage while maintaining the key features of legacy blockchains. The proposed scheme employs the rateless code to encode the block data and store the coded blocks distributedly. Instead of processing the transactions in blocks, as is the case in other existing coded blockchains which may discard the processed blocks when it has a single invalid transaction and hence decreases the throughput, our approach is based on reaching a consensus on each transaction. This fact and an additional transaction assignment stage where a subset of submitted transactions are selected and assigned to miners, as well as low decoding and encoding complexity of raptor codes enable our proposed scheme to achieve higher throughput, distributed storage, and better scalability. Also, we discussed how performing the verification on raw as opposed to coded data (as in Polyshard and LCB) provides a more secure and decentralized scheme closer to the legacy blockchain.\par

\section*{APPENDIX\\RNM ALGORITHM}
In this section, we provide a detailed explanation of the RNM algorithm utilized in our scheme. For a coded block, denoted as
\begin{equation}
 \mathbf{c}_{\ell}(j) = \mathbf{d}_{\ell}\left( i_1 \right) \oplus \mathbf{d}_{\ell}\left( i_2 \right) \oplus \ldots \oplus \mathbf{d}_{\ell}\left( i_{L_j} \right),   
\end{equation}
we define the set of neighbor codeword indices as
\begin{equation}
 \mathcal{N}\left( \mathbf{c}_{\ell}(j) \right) = \left\{ i_1, i_2, \ldots, i_{L_j} \right\}.   
\end{equation}
Additionally, for an intermediate codeword \(\mathbf{d}_{\ell}(i)\), its edge blocks are defined as
\begin{equation}
 \mathcal{E}\left( \mathbf{d}_{\ell}(i) \right) = \left\{ j : i \in \mathcal{N}\left( \mathbf{c}_{\ell}(j) \right) \right\}.   
\end{equation}
Suppose miner \(j\) aims to repair the intermediate codeword \(\mathbf{d}_{\ell}(i)\). By utilizing an edge coded block \(\mathbf{c}_{\ell}\left( j' \right)\), where \(j' \in \mathcal{E}\left( \mathbf{d}_{\ell}(i) \right)\), and the other neighbor intermediate codewords \(\left\{ \mathbf{d}_{\ell}(h) : h \in \mathcal{N}\left( \mathbf{c}_{\ell}\left( j' \right) \right) \setminus \{ i \} \right\}\), miner \(j\) can repair \(\mathbf{d}_{\ell}(i)\) using the following equation:
\begin{equation}\label{yyuujj}
\mathbf{d}_{\ell}(i) = \mathbf{c}_{\ell}\left( j' \right) \oplus \bigoplus_{ h \in \mathcal{N}\left( \mathbf{c}_{\ell}\left( j' \right) \right) \setminus \{ i \} } \mathbf{d}_{\ell}(h), \quad \text{for any } j' \in \mathcal{E}\left( \mathbf{d}_{\ell}(i) \right). 
\end{equation}
However, miner \(j\) does not possess knowledge of the sets of edges \(\mathcal{E}\left( \mathbf{d}_{\ell}(i) \right)\) or neighbors \(\mathcal{N}\left( \mathbf{c}_{\ell}\left( j' \right) \right)\) for any \(j' \in \mathcal{E}\left( \mathbf{d}_{\ell}(i) \right)\). Consequently, miner \(j\) attempts to construct the set \(\hat{\mathcal{E}}\left( \mathbf{d}_{\ell}(i) \right) \subseteq \mathcal{E}\left( \mathbf{d}_{\ell}(i) \right)\). Initially, with an empty set \(\hat{\mathcal{E}}\left( \mathbf{d}_{\ell}(i) \right) = \emptyset\), miner \(j\) broadcasts \(i\) to the network. If \(i \in \mathcal{N}\left( \mathbf{c}_{\ell}(x) \right)\), miner \(x\) responds to miner \(j\) with \(\mathcal{N}\left( \mathbf{c}_{\ell}(x) \right)\), enabling miner \(j\) to include \(\mathbf{c}_{\ell}(x)\) in the set \(\hat{\mathcal{E}}\left( \mathbf{d}_{\ell}(i) \right)\). It's important to note that if a miner \(x\), for which \(\mathbf{c}_{\ell}\left( x \right) \in \mathcal{E}\left( \mathbf{d}_{\ell}(i) \right)\), does not respond to miner \(j\) due to being offline or acting maliciously, \(\mathbf{c}_{\ell}\left( x \right)\) is considered missing and belongs to \(\mathcal{E}\left( \mathbf{d}_{\ell}(i) \right) \setminus \hat{\mathcal{E}}\left( \mathbf{d}_{\ell}(i) \right)\).\par
After obtaining the set \(\hat{\mathcal{E}}\left( \mathbf{d}_{\ell}(i) \right)\), miner \(j\) checks the availability of intermediate codewords \(\left\{ \mathbf{d}_{\ell}(h) : h \in \mathcal{N}\left( \mathbf{c}_{\ell}\left( j' \right) \right) \setminus \{ i \} \right\}\) for each \(\mathbf{c}_{\ell}\left( j' \right) \in \hat{\mathcal{E}}\left( \mathbf{d}_{\ell}(i) \right)\). If there exists a \(j'\) such that all intermediate codewords in \(\mathcal{N}\left( \mathbf{c}_{\ell}\left( j' \right) \right) \setminus \{ i \}\) are available, miner \(j\) can repair the intermediate codeword \(\mathbf{d}_{\ell}(i)\) using Eq. (\ref{yyuujj}). However, if this condition is not satisfied, miner \(j\) resorts to full decoding to obtain the required information.
\end{document}